\documentclass[12pt]{article}
\usepackage{amssymb,amsmath,epsfig,xcolor}
\usepackage{float}
\allowdisplaybreaks

\begin{document}

\title{{\textbf{Physical Existence of Relativistic Stellar Models within
the context of Anisotropic Matter Distribution}}}
\author{M. Sharif$^{1,2}$\thanks{msharif.math@pu.edu.pk}~,
Tayyab Naseer$^1$\thanks{tayyab.naseer@math.uol.edu.pk;
tayyabnaseer48@yahoo.com}~ and Hira Shadab$^1$\thanks{hirashadab123@gmail.com}\\
$^1$Department of Mathematics and Statistics, The University of Lahore,\\
1-KM Defence Road Lahore-54000, Pakistan.\\
$^2$Research Center of Astrophysics and Cosmology, Khazar University, \\
Baku, AZ1096, 41 Mehseti Street, Azerbaijan.}

\date{}
\maketitle

\begin{abstract}
Two distinct non-singular interior models that describe anisotropic
spherical configurations are presented in this work. We develop the
Einstein field equations and the associated mass function in
accordance with a static spherical spacetime. We then discuss
certain requirements that must be satisfied for compact models to be
physically validated. Two distinct limitations are taken into
account to solve the field equations, including different forms of
the radial geometric component and anisotropy, which ultimately
leads to a couple of relativistic models. In both cases, solving the
differential equations result in the appearance of integration
constants. By equating the Schwarzschild exterior metric and
spherical interior line element on the interface, these constants
are explicitly obtained. The disappearance of the radial pressure on
the hypersurface is also used in this context. We further use
estimated radii and masses of six different stars to graphically
visualize the physical properties of new solutions. Both of our
models are deduced to be well-aligned with all physical
requirements, indicating the superiority of the presence of
anisotropy in compact stellar interiors over the perfect isotropic
fluid content.
\end{abstract}

\section{Introduction}

The scientific community regards the general theory of relativity
(GR) as the most accurate theory for understanding the nature of
gravity. The Einstein field equations have been formulated using
fundamental mathematical quantities, particularly the Einstein
tensor and energy-momentum tensor (EMT), to provide a quantitative
description of how matter influences spacetime structure. After the
revolutionary work of Schwarzschild, the subject of finding the
exact solution to equations related to a static sphere in GR and its
extended frameworks has become a fascinating topic. He initially
developed the external \cite{1} and internal solutions \cite{2} in
1916 by taking into account a sphere associated with the constant
density. In order to describe physically feasible stellar objects,
astrophysicists are now interested in finding the solutions of the
gravitational equations that satisfy the necessary physical
constraints. However, because of the non-linearity in the field
equations, it might be challenging to find precise solutions that
describe compact models. Recently, there has been progress in
modeling the physically realistic stars that are associated to
various fluid configurations and spacetime geometries.

In gravitational theories, the field equations are used to fully
describe the structure of self-gravitating objects. These equations
are difficult to solve because the higher derivatives of geometric
components are involved. Depending on the situation, the solution
may be analytic or numerical, and in the later scenario, initial and
boundary conditions are required. Despite that, to get a
comprehensive solution, more information about local physics is
required. In the research of dense bodies, various recently
identified elements revealed to affect the structural
characteristics of such huge objects \cite{56}-\cite{57b}. Finding a
unique solution to the static anisotropic sphere without imposing
constraints, like a well-established formulation of gravitational
potentials, is impossible due to its three independent field
equations with five unknowns (three matter variables and two metric
functions). A well-behaved solution may be found by limiting the
geometric coefficients, which lowers the number of unknowns. The
Karmarkar condition, which provides a differential equation to
derive one metric function in terms of another, has been used by
many scholars \cite{64}-\cite{67}. The case of a disappearing Weyl
tensor represents another key consideration in this framework
\cite{68}.

Stars are the main constituents of our galaxy in the vast and
captivating universe. Thermal reactions take place at the core of
every star, and the outward pressure created by this response
balances the object's gravitational attraction acting inward. Until
the opposing forces are equal in magnitude, the star stays in an
equilibrium state. When gravity takes over and the outward pressure
drops, the stellar object undergoes gravitational collapse. A black
hole, neutron star, or white dwarf might be the outcome of this
phenomenon. It is interesting to note that this classification
depends only on the initial mass of the star. The physical structure
of these huge objects has captivated astronomers, who have focused
on studying their evolution. Neutron stars, which are produced when
large stellar masses collapse due to gravity, were the most
interesting of all the other compact candidates. Their mass lies
between 1-3 times that of our sun, and their dense core is composed
of freshly generated neutrons. The star would not collapse again
because of the pressure created by these neutrons, which resists the
pull of gravity. Despite the fact that the concept of a neutron star
initially came across in 1934 \cite{79}, a considerable time was
spent for its demonstrative proof because these structures usually
avoid from being directly detected because they emit insufficient
radiation.

The ground breaking research of Jeans \cite{3} revealed that several
parameters controlling the internal geometry indicated the nature of
fluid to be anisotropic in nature. Shortly after this, Lemaitre
\cite{4} introduced the initial anisotropic solution with unchanging
density and pressure confined to radial and tangential components.
Theoretically, Ruderman \cite{5} found that huge star interiors
(with densities more than $10^{15}g/cm^3$) might not have radial and
tangential pressures to be equal. Numerous factors like the
existence of pion condensation \cite{10}, powerful magnetic fields
\cite{6}-\cite{9}, nuclear forces \cite{11}, and superfluid region
\cite{12} etc. have been identified that produce anisotropy within a
dense body. Initially, Bowers and Liang \cite{13} examined how the
anisotropy affected the stability of large objects and proposed that
anisotropy should not be present in the core. They claimed that the
equilibrium mass of the star and its surface redshift are
significantly impacted by this factor. Numerous observations have
been made in order to examine how anisotropy affects several
characteristics of stellar models, including mass, radius,
non-radial oscillation, and moment of inertia \cite{14}-\cite{18}.
It was discovered that the magnitudes of the above described
features directly correlate with that of the pressure anisotropy.

Lake \cite{19} used the Newtonian equation to develop the
corresponding anisotropic analog of the isotropic spherical
spacetime. Herrera and Santos \cite{20} investigated both Newtonian
and relativistic fluids, and examined the potential causes of
anisotropy in self-gravitating objects. They provided some
relationships with the Weyl and shear tensors in order to better
investigate its function. Herrera \cite{21} used pressure anisotropy
while studying relativistic fluid distributions. He demonstrated
that during compact evolutionary phases, physical processes can lead
to anisotropy, even if the fluid is initially isotropic. It is vital
to highlight that any equilibrium configuration depicts the last
phase of a dynamic process and there is no proof that anisotropy
will vanish in the final state of equilibrium. Hence, even when
starting from an isotropic pressure state, the end state
configuration will theoretically display anisotropic pressure
features. The presence of anisotropy inside giant stars is therefore
not an exception, rather it necessarily exists. Numerous scholars
have looked at how pressure anisotropy affects relativistic dense
stats and have discovered results that are physically relevant
\cite{47}-cite{40}. Sharif and his associates \cite{37,38} creates
anisotropic extensions of various isotropic spacetimes and explored
the stability under particular parametric ranges.

The anisotropic solutions admitting spherical symmetry are developed
in this research. The format of this paper is provided as follows.
In the next section, we formulate the Einstein field equations
assuming a spherical static spacetime associated with anisotropic
matter content. A well-behaved solution must meet certain physical
conditions which are listed in section \textbf{3}, so that we can
analyze them for our novel solutions. In section \textbf{4}, we
model a couple of anisotropic solutions using certain radial metric
potentials and two anisotropies. The constants involved in metric
components are obtained by applying the boundary conditions between
the Schwarzschild exterior and the considered interior line
elements. Additionally, a more-detailed graphical examination of the
two constructed solutions is also given. The significance and
implications of our results are finally presented in section
\textbf{5}.

\section{Static Sphere and Einstein Field Equations}

The line element is an essential mathematical tool for expressing
the internal spacetime of a celestial object. It is also important
to understand the behavior of energy and matter inside these
self-gravitating bodies. One of the primary advantages of the static
spherical line element is the simplification of the field equations.
In order to initiate our study, we suppose a spherical interior
metric as
\begin{equation}\label{g6}
ds^2=-\mathrm{A}_1^2 dt^2+\mathrm{B}_1^2 dr^2+r^2 d\Omega^2,
\end{equation}
where $\mathrm{A}_1=\mathrm{A}_1(r),~\mathrm{B}_1=\mathrm{B}_1(r)$
and $d\Omega^2=d\theta^2+\sin^2\theta d\phi^2$.

The Einstein-Hilbert action is the foundational principle that
elegantly encodes the dynamics of spacetime curvature in relation to
matter and energy. In the framework of classical field theory,
physical laws are often derived by extremizing an action functional.
For the gravitational theories, the action must satisfy two key
requirements: \emph{(i}) it should be invariant under arbitrary
coordinate transformations, and \emph{(ii)} it should be the
simplest possible scalar invariant (the Ricci scalar $R$) involving
the curvature of spacetime. The action functional, involving the
matter Lagrangian density $L_m$, is defined as follows
\begin{equation}\label{g2a}
I=\int\sqrt{-g}\left(\frac{R}{16\pi}+L_{m}\right)d^{4}x,
\end{equation}
from which the Einstein field equations can be obtained using the
variational principle as
\begin{equation}\label{g2}
G_{\alpha\eta}=8\pi T_{\alpha\eta},
\end{equation}
where $G_{\alpha\eta}=R_{\alpha\eta}-\frac{1}{2}R g_{\alpha\eta}$
with $R_{\alpha\eta}$, and $g_{\alpha\eta}$ being the Ricci tensor,
and metric tensor, respectively. The geometrical structure is
represented by $G_{\alpha\eta}$, also referred to the Einstein
tensor. Moreover, the value on the right hand side,
$T_{\alpha\eta}$, represents the usual matter source.

The anisotropic fluid interiors have attracted a lot of interest
from scientists studying a variety of physical phenomena. Unlike
perfect isotropic matter configuration where the physical factors
remain the same in each direction, anisotropic fluid capture the
crucial orientation-based variations in these quantities. This
capability becomes significant particularly when analyzing systems
exhibiting intrinsic directional symmetry, such as relativistic
compact objects or extreme gravitational environments. Given our
focus on anisotropic compact configurations, the EMT describing such
an interior matter distribution is given by \cite{fag}-\cite{fak}
\begin{equation}\label{g5}
T_{\alpha\eta}=(p_t+\rho)\mathrm{U}_{\alpha}\mathrm{U}_{\eta}-p_t
g_{\alpha\eta}-(p_t-p_r)\mathrm{V}_\alpha\mathrm{V}_\eta,
\end{equation}
where $p_r,~p_t,~\rho,~\mathrm{U}_{\alpha}$ and
$\mathrm{V}_{\alpha}$ represent the radial pressure, tangential
pressure, energy density, four-velocity and four-vector,
respectively. In the co-moving frame of reference, the final two
physical terms become under the line element \eqref{g6} as
\begin{equation}\nonumber
\mathrm{U}_\alpha=(\mathrm{-A}_1,0,0,0), \quad
\mathrm{V}_\alpha=(0,\mathrm{B}_1,0,0),
\end{equation}
satisfying the following requirements
$$\mathrm{U}^\alpha \mathrm{U}_{\alpha}=-1, \quad \mathrm{V}^\alpha
\mathrm{V}_{\alpha}=1, \quad \mathrm{U}^\alpha
\mathrm{V}_{\alpha}=0.$$ The non-vanishing components of EMT
\eqref{g5} are
\begin{eqnarray}\nonumber
&T_{00}=\rho\mathrm{A}_1^2, \quad T_{11}=p_r\mathrm{B}_1^2,\quad
T_{22}=p_tr^2,\quad T_{33}=p_tr^2\sin^2\theta.
\end{eqnarray}

The independent components of the field equations \eqref{g2} are
determined under the spherical metric \eqref{g6} and the fluid
\eqref{g5} as described below
\begin{align}\label{g8}
8\pi\rho&=-\frac{1}{r^2\mathrm{B}_1^2}+\frac{2\mathrm{B}_1'(r)}{r\mathrm{B}_1(r){}^3}+\frac{1}{r^2},
\\\label{g9}
8\pi{p}_r&=\frac{2\mathrm{A}_1'}{r\mathrm{A}_1
\mathrm{B}_1^2}+\frac{1}{r^2 \mathrm{B}_1^2}-\frac{1}{r^2},
\\\label{g10}
8\pi{p}_t&=-\frac{\mathrm{A}_1'\mathrm{B}_1'}{\mathrm{A}_1
\mathrm{B}_1^3}+\frac{\mathrm{A}_1'}{r\mathrm{A}_1
\mathrm{B}_1^2}+\frac{\mathrm{A}_1''}{\mathrm{A}_1
\mathrm{B}_1^2}-\frac{\mathrm{B}_1'}{r\mathrm{B}_1^3},
\end{align}
where prime represents the derivatives with respect to $r$. In
relativistic astrophysics, the mass function quantifies the
gravitational mass enclosed within a given radius of a spherically
symmetric spacetime, playing a key role in modeling compact objects.
The expression for the Misner-Sharp mass $m(r)$ is \cite{faj}
\begin{align}\label{g12a}
m(r)=4\pi\int_0^r\rho r^2 dr \quad \Longrightarrow \quad m'(r)=4\pi
r^2\rho(r).
\end{align}
The above differential equation can be resolved either numerically
or analytically, depending on how complicated the expression for the
energy density is. Equations \eqref{g9} and \eqref{g10} can  be used
to express the anisotropy of the system as
\begin{align}\label{g12d}
\mathrm\Pi(r)=8\pi(p_t-p_r)=-\frac{\mathrm{A}_1'\mathrm{B}_1'}{\mathrm{A}_1
\mathrm{B}_1^3}+\frac{\mathrm{A}_1'}{r\mathrm{A}_1
\mathrm{B}_1^2}+\frac{\mathrm{A}_1''}{\mathrm{A}_1
\mathrm{B}_1^2}-\frac{\mathrm{B}_1'}{r\mathrm{B}_1^3}.
\end{align}
At the interior of the stellar configuration, $\mathrm\Pi(r)$ should
vanish, meaning that $p_r=p_t$ at that point. A positive anisotropic
force induces repulsion that affect the stability by increasing
pressure and density at the center. On the other hand, a negative
force can cause contraction, decreasing pressure and density of the
object.
\section{Physical Acceptability Criteria for Relativistic Models}

The Einstein gravitational equations that describe the interior
composition of dense objects can be solved analytically or
numerically using a variety of methods that have been proposed in
the literature. Several authors compiled physically acceptable
conditions, the fulfilment of which guarantees the physical
viability of resulting model \cite{ab,ac}. In the following, we
address these physical conditions.
\begin{itemize}
\item The presence of geometric singularities within the star is one of
the basic components to be avoided. In order to verify that the
metric potentials are not singular within a dense structure, we must
examine the behavior of both time and radial functions
mathematically as well as graphically. In a feasible
self-gravitating interior, the behavior of these components should
be finite in the core and positively rising outwards.

\item The pattern of matter variables, including pressure
components and energy density, should be positive and finite
everywhere. Additionally, they have to attain their minimum
(maximum) at $\Sigma:r=R$ ($r=0$). Similarly, it must be ensured
that these variables will trend downward towards the boundary by
checking if their first order derivatives vanish at $r=0$ and become
negative, otherwise.

\item It should be noted that the existence of anisotropy, regardless of its sign,
has a major impact on the self-gravitating stability of the model.
An important factor in improving the stability of the star is its
positive nature, which is defined by $p_r$ being lower than $p_t$.
This is explained by the pressure that is directed outward, which
prevents the star from collapsing by acting as a supporting force
that opposes gravity. On the other hand, the star may become
unstable if there is negative anisotropy, in which $p_r$ exceeds
$p_t$ since there is no outward directed pressure. Therefore,
compared to a compact body having positive anisotropy, the stars
admitting its negative profile are usually stable for less duration
of time.

\item The compactness, or mass-radius ratio, describes how firmly
particles are bonded to one another in a self-gravitating system.
The gravitational pull of extremely dense bodies, like neutron
stars, is very strong. Its mathematical form is given as
\begin{equation}\label{g40}
\gamma(r)=\frac{m(r)}{r}.
\end{equation}
In a spherical interior, its value must be $<0.44$ everywhere
\cite{42a}. Later research has shown that this limit can also be
applied to anisotropic matter distributions, even though it was
firstly suggested for an isotropic fluid sphere. Furthermore, a
phenomenon, named the surface redshift, where light emitted from a
massive object loses energy while escaping a strong gravitational
field, causing its wavelength to stretch and shift toward red part
of the spectrum. This happens because gravity drains energy from the
light. The more intense the gravity, the more extreme the redshift
will be. The amount of matter contained within a star and its
compactness are used to determine the surface redshift. This is
provided as
\begin{equation}\label{g41}
z(r)=\bigg(1-\frac{2m}{r}\bigg)^\frac{-1}{2}-1.
\end{equation}
As the anisotropic matter content is under investigation in this
analysis, at the surface, the redshift reaches its peak value by
$5.211$ in order to obtain an acceptable model \cite{42b}.

\item An expression for the equation of state is $p_j = \omega_j \rho$,
where $\omega_j$ with $j=r,t$ stand for radial and transverse
elements, respectively. They can be written as follows
\begin{equation}\label{gg41}
\omega_r=\frac{p_r}{\rho}, \quad \omega_t=\frac{p_t}{\rho}.
\end{equation}
A physically feasible fluid arrangement requires that the ratios of
energy density to pressure components fall within a certain range of
$[0,1]$.

\item Another essential element for confirming the accuracy
of a stellar object are the energy conditions which are defined as
the linear combinations of the fluid variables. To ensure that the
dense object involves usual/ordinary fluid, they must be positive
inside a star. For the current configuration, these energy
conditions are

$Null: \quad \rho+p_t\geq0, \quad \rho+p_r\geq0$,

$Weak: \quad \rho+p_t\geq0, \quad \rho+p_r\geq0, \quad \rho\geq0$,

$Dominant: \quad \rho\pm p_t\geq0, \quad \rho\pm p_r\geq0$,

$Strong: \quad \rho+p_r+2p_t\geq0, \quad \rho+p_t\geq0, \quad
\rho+p_r\geq0$,

$Trace: \quad \rho-p_r-2p_t\geq0$.

As both pressures and density are positive, this ultimately results
in positive weak, strong and null energy conditions, so we just need
to study dominant and trace bounds.

\item The hydrostatic equilibrium of the system is lost as a result of
variations in the internal configuration caused by several
components. Anisotropic, hydrostatic, and gravitational forces are
the three candidates that are used to verify the equilibrium state
of any compact model. According to the criterion of validity, the
net sum of all the associated forces become null, mathematically
represented by
\begin{eqnarray}\label{hg52}
f_g+f_h+f_a=0.
\end{eqnarray}
In the presence of anisotropy, the internal geometry of spherically
symmetric dense celestial objects is expressed by the following
evolution equation as \cite{42c,42d}
\begin{eqnarray}\label{hhg52}
-\frac{\mathrm{M}_G}{r}(\rho+p_r)\frac{\mathrm{A}_1}{\mathrm{B}_1}-\frac{2}{r}(p_r-p_t)-
p_r'=0,
\end{eqnarray}
where $\mathrm{M}_G(r)$ being the gravitational mass whose value is
\begin{eqnarray}\label{hgg52}
\mathrm{M}_G(r)=\frac{r\mathrm{B}_1\mathrm{A}_1'}{\mathrm{A}_1^2},
\end{eqnarray}
whose substitution in Eq.\eqref{hhg52} yields
\begin{eqnarray}\label{hg53}
-\frac{\mathrm{A_1'}}{A_1}(\rho+p_r)-\frac{2}{r}(p_r-p_t)- p_r'=0.
\end{eqnarray}
The combination of Eqs.\eqref{hg52} and \eqref{hg53} results in
different forces as
\begin{eqnarray}
&f_a&=\frac{2}{r}(p_t-p_r),\\ \label{hhh52} &f_h&=-p_r',\\
\label{ii52} &f_g&=-\frac{\mathrm{A}_1'}{\mathrm{A}_1}(\rho+p_r).
\end{eqnarray}

\item Examining the structural stability of celestial systems has become a
compelling subject in astrophysical studies. One of the various
phenomena proposed in the literature is based on the speed of sound,
with tangential and radial components given by
$v_{r}^{2}=\frac{dp_{r}}{d\rho}$ and
$v_{t}^{2}=\frac{dp_{t}}{d\rho}$, respectively. According to Abreu
et al. \cite{42bb}, causality would be maintained in a system if the
speed of light is greater than the speed of sound, i.e., $0 <
v_{r}^{2}, v_{t}^{2} < 1$.

Similarly, Herrera \cite{42ba} suggested that cracking would take
place in the internal fluid if a change in the sign of a radial
force occurs during the evolution. To get a stable model, it must be
avoided. He explained that the cracking would not occur only when $0
< |v_{t}^{2}-v_{r}^{2}| < 1$ holds.

An important factor in determining the stability of compact stars is
the adiabatic index, which is represented by $\Gamma$. The specific
heat at constant volume divided by the specific heat at constant
pressure is represented by this dimensionless parameter. This is
also a measurement of the inner interaction and heat behavior of
dense objects. Astrophysicists can learn more about the consistency
requirements, pressure-density interactions, and general stability
of the compact stellar objects by looking at this factor. The
adiabatic index, in the case of anisotropic fluid, is expressed
mathematically as
$$\Gamma_{r}=\frac{p_{r}+\rho}{p_{r}}\frac{dp_{r}}{d\rho}, \quad
\Gamma_{t}=\frac{p_{t}+\rho}{p_{t}}\frac{dp_{t}}{d\rho}.$$
Importantly, for a stable structure, the factors listed above must
always be bigger than $\frac{4}{3}$ \cite{42f}.
\end{itemize}

\section{Novel Analytical Solutions under Anisotropic Fluid Source}

Researchers have devised multiple analytical and numerical
approaches to solve the Einstein field equations, motivated by
persistent theoretical interest. We must now focus our efforts on
overcoming this hurdle through a novel methodology. While various
theoretical frameworks might resolve this problem, we direct our
attention to two particularly insightful anisotropic factors along
with $g_{rr}$ metric functions. By selecting two different
constraints, we balance the number of equations with the number of
unknowns in the next subsections to find solutions of such
non-linear differential equations.

\subsection{Model I}

After carefully examining various metric functions in the
literature, researchers are encouraged to take into account known
forms of the temporal and radial geometric terms. To continue our
analysis, we thus select a specific $g_{rr}$ metric potential which
is provided as \cite{88a,88b}
\begin{align}\label{g52}
\mathrm{B}_1^2(r)=\frac{a+2br^2}{a+br^2},
\end{align}
with $a$ (dimensionless) and $b$ (having dimension of
$\frac{1}{\ell^2}$) being arbitrary constants. This particular
metric, applied in the study of anisotropic interior compact models,
maintains a finite and continuous behavior throughout. Moreover,
$\mathrm{B}_1^2(r)|_{r=0}=1$ and
$\big(\mathrm{B}_1^2(r)\big)'|_{r=0}=0$ ensure that the associated
metric potential is both regular and singularity-free as shown in
Figure \textbf{1}. By substituting the radial metric component
\eqref{g52} in Eq.\eqref{g12d}, the following expression for
anisotropy can be derived
\begin{align}\label{g52a}
\Pi(r)&=\frac{-\big(a^2+4 a b r^2+2 b^2 r^4\big)\mathrm{A}_1'+r
\big(a^2+3 a br^2+2 b^2 r^4\big)\mathrm{A}_1''+2 b^2 r^3
\mathrm{A}_1}{8 \pi r\mathrm{A}_1 \big(a+2 b r^2\big)^2},
\end{align}
which is a differential equation of second order. We modify the
above equation as follows
\begin{align}\label{g53i}
\frac{\mathrm{A}_1''}{\mathrm{A}_1}-\frac{\mathrm{A}_1'}{\mathrm{A}_1}
\bigg[\frac{a^2+4abr^2+2b^2r^4}{r\big(a^2+3abr^2+2b^2r^4\big)}\bigg]=
\frac{r\Pi(r)\big(a+2br^2\big)^2-2b^2r^3}{r\big(a^2+3abr^2+2b^2r^4\big)}.
\end{align}

We can solve Eq.\eqref{g53i} to find $\mathrm{A}_1(r)$ by specifying
a unique expression of the factor $\Pi(r)$. We need to configure
this element so that it effectively cancels out at the geometric
center of astrophysical object. Moreover, the behavior of this
quantity must exhibit a monotonically rising trend (i.e., $p_t>p_r$)
with increasing radial distance to support the formation of stellar
structures \cite{47}. In this work, we select a form of anisotropy
that meets this criteria and simplifies the process of solving the
Eq.\eqref{g53i}, given by
\begin{align}\label{g53c}
\Pi(r)=\frac{2b^2r^2}{(a+2br^2)^2},
\end{align}
which leads Eq.\eqref{g53i} to
\begin{align}\label{g53d}
\frac{\mathrm{A}_1''}{\mathrm{A}_1}-\frac{\mathrm{A}_1'}{\mathrm{A}_1}
\bigg[\frac{a^2+4abr^2+2b^2r^4}{r\big(a^2+3abr^2+2b^2r^4\big)}\bigg]=0.
\end{align}
The solution of Eq.\eqref{g53d} is
\begin{align}\label{g53h}
\mathrm{A}_1=\frac{C_1 \left\{4 \mathrm{f}_1-\sqrt{2} a \ln
\left(2\sqrt{2} \mathrm{f}_1+\mathrm{f}_2\right)\right\}}{8b}+C_2,
\end{align}
where $\mathrm{f}_1=\sqrt{a+br^2}\sqrt{a+2 b r^2}$ and $f_2=3 a+4b
r^2$. The constants $C_1$ and $C_2$ will be calculated later. We
note that the given metric function exhibits regular behavior, as it
fulfills the condition of being a $\mathrm{constant}$ at $r=0$ and
$\big(\mathrm{A}_1^2(r)\big)'|_{r=0}=0$. Its physically suitable
nature is further demonstrated in Figure \textbf{1}. Consequently,
the matter components $(\rho,p_r,p_t)$, based on the specific forms
of the metric functions \eqref{g52} and \eqref{g53h}, are expressed
as follows
\begin{align}\label{g54i}
8\pi\rho&=\frac{b(3a+2br^2)}{(a+2br^2)^2},
\\\label{g55i}
8\pi p_r&=\frac{16bC_1 \sqrt{a+b r^2}}{\sqrt{a+2 b
r^2}(C_1(4\mathrm{f}_1-\sqrt{2} a \ln
(2\sqrt{2}\mathrm{f}_1+\mathrm{f}_2))+8 bC_2)}-\frac{b}{a+2 b r^2},
\\\nonumber 8\pi p_t&=
\frac{b}{\mathrm{f}_1^{5/2}(2 \sqrt{2}\mathrm{f_1}+
(\mathrm{f_2}^2))(C_1(4 \mathrm{f_1}-\sqrt{2}a\ln(2
\sqrt{2}\mathrm{f_1}+\mathrm{f_2}))+8bC_2)}
\\\nonumber &\times\big[
C_1\{a^2(24 a^3+a^2(17 \sqrt{2}\mathrm{f}_1+104 b r^2)+32 b^2
r^4(\sqrt{2}\mathrm{f}_1+2 b r^2)(48 a b r^2\\\nonumber&\times
(\sqrt{2}\mathrm{f}_1+3 b r^2))\ln(2
\sqrt{2}\mathrm{f}_1+\mathrm{f}_2+4(3 a^3+17 a^2 b r^2+30 a b^2
r^4+16 b^3r^6))\\\nonumber&\times((17 a^2+12 a
(\sqrt{2}\mathrm{f}_1+4br^2)16 b r^2 (\sqrt{2}\mathrm{f}_1+2 b
r^2))))\}-8 a b C_2 \{12 \sqrt{2} a^3\\\label{g56h} &+a^2(17
\mathrm{f}_1+52 \sqrt{2} b r^2)+32 b^2 r^4 \mathrm{f}_1+\sqrt{2} b
r^2\}+24 a b r^2(2\mathrm{f}_1+3 \sqrt{2} b r^2)\big].
\end{align}
\begin{figure}\center
\epsfig{file=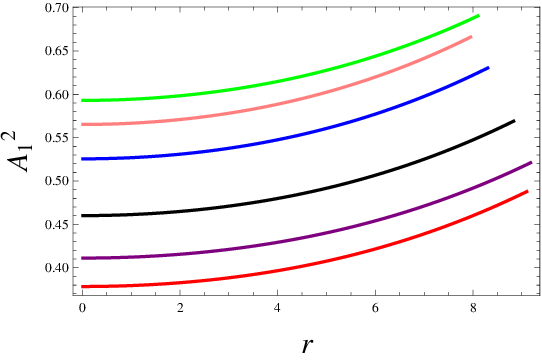,width=0.43\linewidth}\epsfig{file=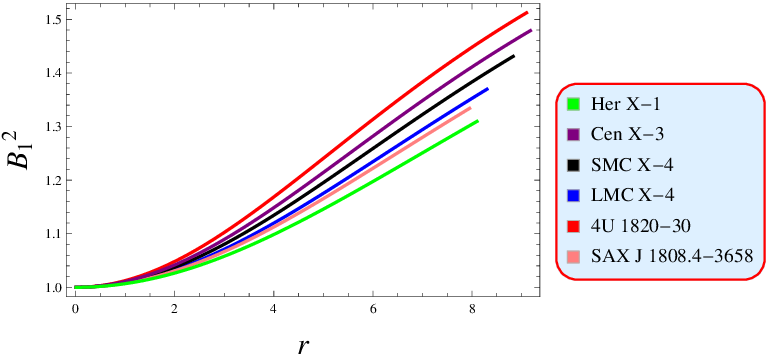,width=0.6\linewidth}
\caption{Temporal \eqref{g53h} and radial \eqref{g52} metric
functions for $b=0.02$ corresponding to model I.}
\end{figure}

The matching conditions governing field behavior and their
differential properties across matter interfaces or spatial
boundaries play a fundamental role in compact star's modeling. These
requirements, including metric continuity and curvature matching,
guarantee smooth spacetime transitions and maintain field equations'
validity. The Schwarzschild exterior metric represents a distinctive
solution to the gravitational field equations for a static,
spherically symmetric spacetime within the context of a vacuum.
Therefore, this is the appropriate line element describing the outer
geometry, expressed as
\begin{equation}\label{g15}
ds^2=-\bigg(1-\frac{2\mathrm{M}}{r}\bigg)dt^2
+\bigg(1-\frac{2\mathrm{M}}{r}\bigg)^{-1}dr^2+r^2d\Omega^2.
\end{equation}
Here, the total mass of the exterior region is denoted by
$\mathrm{M}$, with the condition that $r>2\mathrm{M}$. In model I,
the constants $(a,C_1,C_2)$ are already introduced which can be
calculated using matching conditions. They ensure continuous
differentiability of curvature tensors (both intrinsic and
extrinsic) at the boundary $(\Sigma:r=R)$.

The first fundamental form enforces continuity in the temporal and
radial components of the metric, specifically, $g_{tt}$ and $g_{rr}$
for the metrics \eqref{g6} and \eqref{g15}. This leads to
\begin{align}\label{g16}
\mathrm{A}_1^2(R)&~=~1-\frac{2\mathrm{M}}{R},
\\\label{g17}
\mathrm{B}_1^2(R)&~=~\bigg(1-\frac{2\mathrm{M}}{R}\bigg)^{-1},
\end{align}
which yield after joining with Eqs.\eqref{g52} and \eqref{g53h} as
follows
\begin{align}\label{g16a}
\frac{C_1 \left\{4\mathrm{f}_1^*-\sqrt{2} a \ln \left(2 \sqrt{2}
\mathrm{f}_1^*+\mathrm{f}_2^*\right)\right\}}{8
b}+C_2&~=~\sqrt{1-\frac{2\mathrm{M}}{R}},\\\label{g17a}
\frac{a+bR^2}{a+2bR^2}&~=~1-\frac{2\mathrm{M}}{R},
\end{align}
where $\mathrm{f}_1^*=\mathrm{f}_1(R)$ and
$\mathrm{f}_2^*=\mathrm{f}_2(R)$. Equations \eqref{g16a} and
\eqref{g17a} involve three unknown variables, necessitating an
additional constraint for a unique solution. It is widely recognized
in existing studies that the radial pressure vanishes at a certain
finite $r$, which corresponds to the boundary of the star, denoted
here by $R$. This condition, when coupled with Eq.\eqref{g55i},
gives
\begin{align}\label{g18}
p_r(R)&=\frac{16C_1 \sqrt{a+b R^2}}{\sqrt{a+2b
R^2}(C_1(4\mathrm{f}_1^*-\sqrt{2}a\ln
(2\sqrt{2}\mathrm{f}_1^*+\mathrm{f}_2^*))+8bC_2)}-\frac{1}{a+2
bR^2}=0.
\end{align}
After combining Eqs.\eqref{g16a}-\eqref{g18}, the following
constants are obtained as
\begin{align}\label{gg18}
a&=\frac{bR^3-4 b\mathrm{M}R^2}{2\mathrm{M}},
\\\label{ggg18}
C_1&=\frac{b}{\mathrm{f}_1^*}\sqrt{1-\frac{2\mathrm{M}}{R}},\\\label{gggg18}
C_2&=\frac{1}{16}\sqrt{1-\frac{2\mathrm{M}}{R}}
\bigg(\frac{\sqrt{2}a\ln \left(2
\sqrt{2}\mathrm{f}_1^*+\mathrm{f}_2^*\right)}{\mathrm{f}_1^*}+12\bigg).
\end{align}
\begin{figure}\center
\epsfig{file=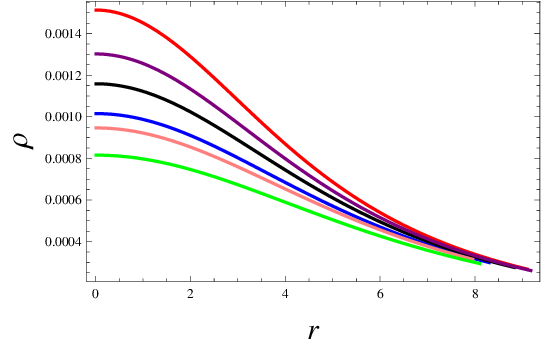,width=0.423\linewidth}\epsfig{file=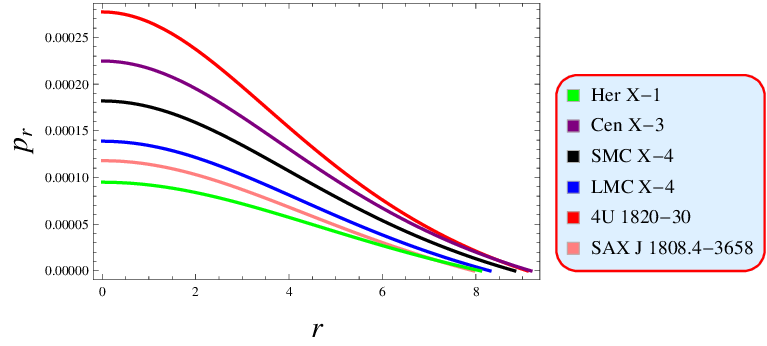,width=0.61\linewidth}
\epsfig{file=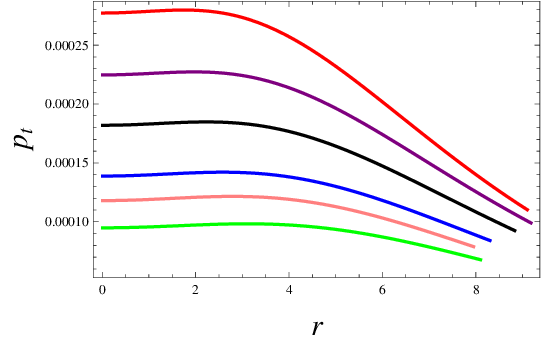,width=0.43\linewidth}\epsfig{file=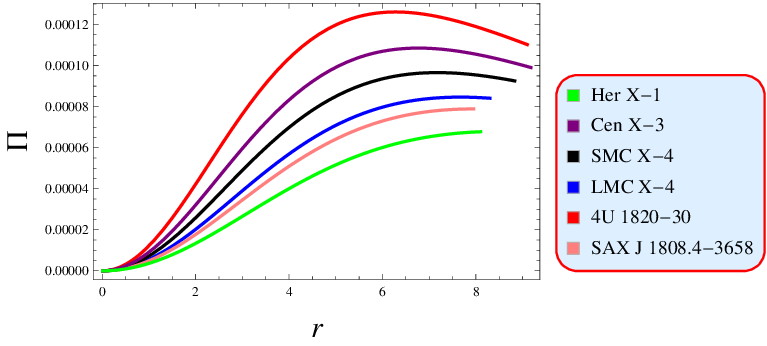,width=0.61\linewidth}
\caption{Fluid parameters for $b=0.02$ corresponding to model I.}
\end{figure}
\begin{table}
\scriptsize \centering \caption{Masses and radii of six different
stars.} \label{Table2} \vspace{+0.1in}
\setlength{\tabcolsep}{1.55em}
\begin{tabular}{cccccc}
\hline\hline \textbf{Stars} &
$\textbf{Mass}~(\mathrm{M}_{\bigodot})$ & $\textbf{Radius (km)}$
\\\hline $\mathrm{Her X-1}$ & $0.85$ & $8.1$
\\\hline $\mathrm{Cen X-3}$ & $1.49$ & $9.178$
\\\hline $\mathrm{SMC X-4}$ & $1.29$ & $8.83$
\\\hline $\mathrm{LMC X-4}$ & $1.04$ & $8.301$
\\\hline $\mathrm{4U 1820-30}$ & $1.58$ & $9.1$
\\\hline $\mathrm{SAX J 1808.4-3658}$ & $0.9$ & $7.951$\\
\hline\hline
\end{tabular}
\end{table}

We use the predicted radii and masses from various star models, as
shown in Table \textbf{1}, to examine the physical behavior of the
resulting solutions. It is observed that both energy density and
pressure attain their highest values at the center of the resulting
configuration (Figure \textbf{2}). At the surface, the pressure
component in the radial direction nullifies, while at the core, both
pressures are equal, (i.e., $p_t(0)=p_r(0)$), resulting in zero
anisotropy at that point (as shown in the lower right panel).
Additionally, the anisotropy increases with radius, supporting the
stability and construction of dense stellar structures. Table
\textbf{2} presents the corresponding analytical values of these
quantities on the basis of which we can claim that our model I
supports the existence of ultra-dense stellar configuration.
Further, Figure \textbf{3} confirms the regular nature of the matter
profiles by illustrating the required behavior of their first and
second derivatives, mathematically, given by
$\frac{d\rho}{dr}|_{r=0}=0=\frac{dp_r}{dr}|_{r=0}=\frac{dp_t}{dr}|_{r=0}$
and
$\frac{d^2\rho}{dr^2}|_{r=0}<0,~\frac{d^2p_r}{dr^2}|_{r=0}<0,~\frac{d^2p_t}{dr^2}|_{r=0}<0$.
\begin{figure}\center
\epsfig{file=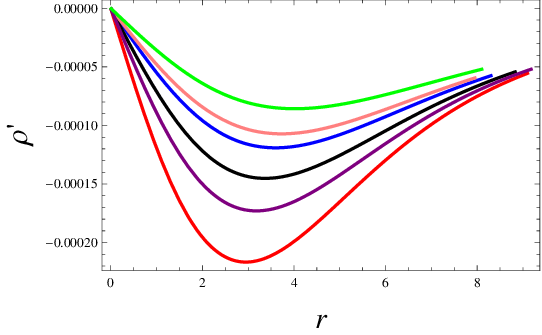,width=0.426\linewidth}\epsfig{file=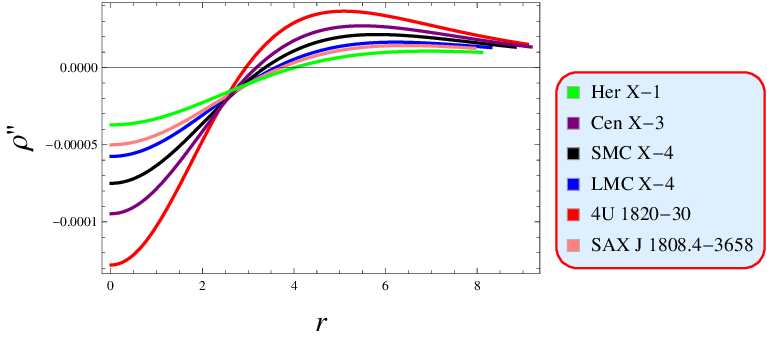,width=0.6\linewidth}
\epsfig{file=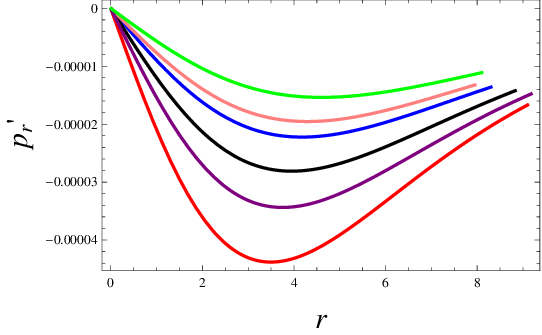,width=0.426\linewidth}\epsfig{file=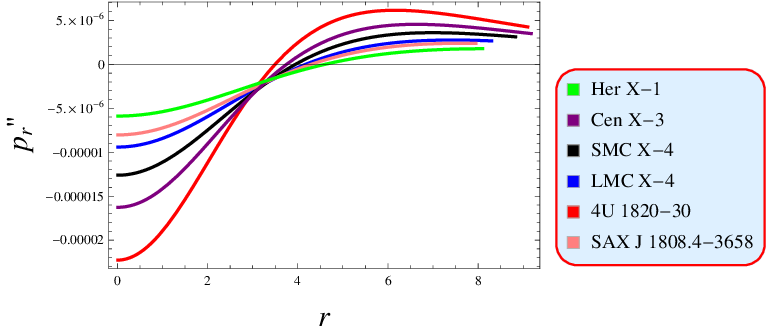,width=0.6\linewidth}
\epsfig{file=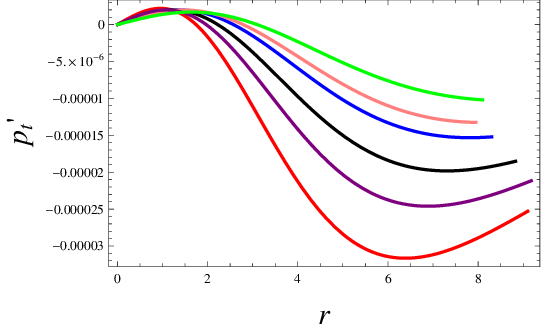,width=0.426\linewidth}\epsfig{file=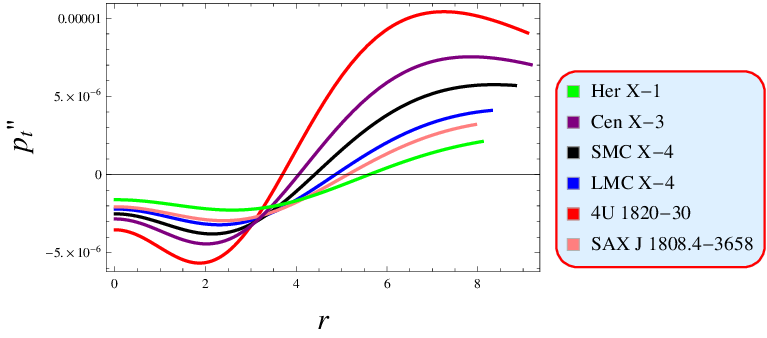,width=0.6\linewidth}
\caption{Regularity conditions for $b=0.02$ corresponding to model
I.}
\end{figure}
\begin{table}
\scriptsize \centering \caption{Various physical factors for
different stars corresponding to model I.} \label{Table2}
\vspace{+0.1in} \setlength{\tabcolsep}{1.55em}
\begin{tabular}{cccccc}
\hline\hline \textbf{Stars} & $\rho_c {\bf (gm/cm^3)}$ & $\rho_s
{\bf (gm/cm^3)}$ & $p_{c} {\bf (dyne/cm^2)}$
\\\hline $\mathrm{Her X-1}$ & 1.0910$\times$10$^{15}$ & 3.8583$\times$10$^{14}$ &
1.1402$\times$10$^{35}$
\\\hline
$\mathrm{Cen X-3}$ & 1.7312$\times$10$^{15}$ &
3.4262$\times$10$^{14}$ & 2.6813$\times$10$^{35}$
\\\hline
$\mathrm{SMC X-4}$ & 1.5412$\times$10$^{15}$ &
3.6242$\times$10$^{14}$ & 2.1788$\times$10$^{35}$
\\\hline $\mathrm{LMC X-4}$ & 1.3525$\times$10$^{15}$ & 3.9212$\times$10$^{14}$ &
1.6629$\times$10$^{35}$
\\\hline
$\mathrm{4U 1820-30}$ & 2.0175$\times$10$^{15}$ &
3.5065$\times$10$^{14}$ & 3.3319$\times$10$^{35}$
\\\hline
$\mathrm{SAX J 1808.4-3658}$ & 6.5755$\times$10$^{15}$ &
4.1968$\times$10$^{14}$ &
1.4128$\times$10$^{35}$ \\
\hline\hline
\end{tabular}
\end{table}
\begin{figure}\center
\epsfig{file=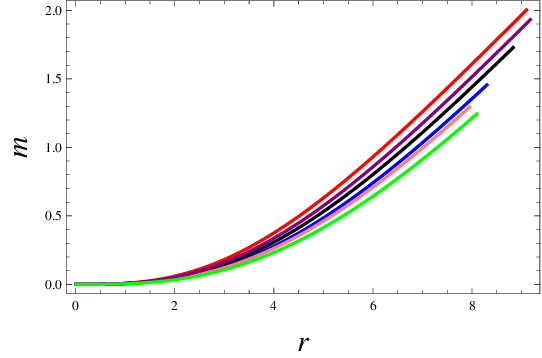,width=0.424\linewidth}\epsfig{file=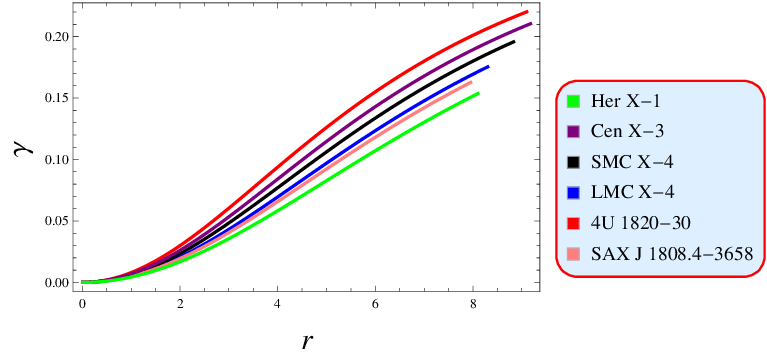,width=0.6\linewidth}
\epsfig{file=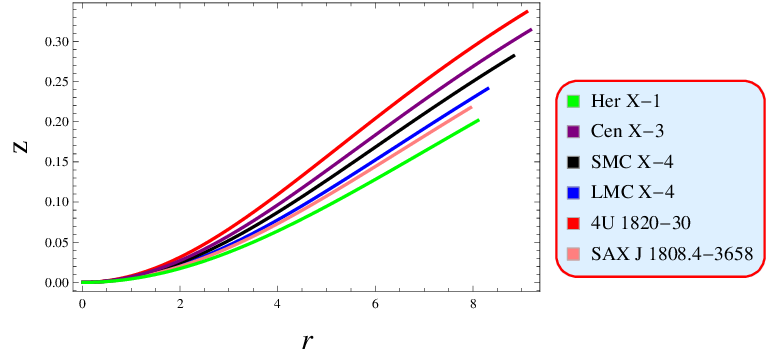,width=0.6\linewidth} \caption{Physical
factors for $b=0.02$ corresponding to model I.}
\end{figure}

A continuous growing pattern of mass function is displayed in Figure
\textbf{4} (top left). The compactness and redshift in the other two
graphs similarly exhibit the behavior consistent with the observed
data. Figure \textbf{5} depicts the behavior of equation of state
parameters, and shows feasibility. Thus, the obtained solution is
proved to be physically acceptable for stellar objects under
investigation. Figure \textbf{6} shows trace and dominant energy
conditions, with positive behavior everywhere in $0<r<R$. As a
result, our solution admits physical viability and contains the
ordinary fluid in the interior distribution. The fundamental forces
are graphically checked in Figure \textbf{7} (top left), and we find
their net sum to be zero, claiming our model in the equilibrium
state for all compact stars. Further, using the cracking and sound
speed criteria, we examine the stability in the same Figure, and it
is confirmed that our model I is stable. The stability conditions
demonstrated in Figure \textbf{8} receive an additional validation
of our model stability through the analysis of an adiabatic index.
\begin{figure}\center
\epsfig{file=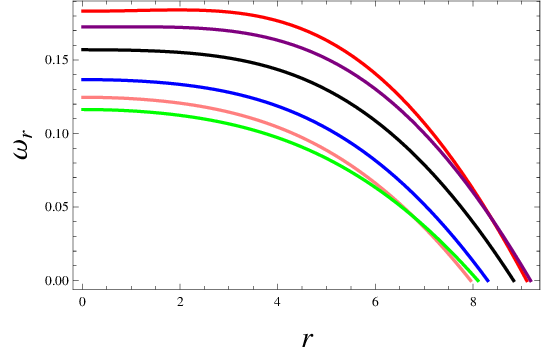,width=0.43\linewidth}\epsfig{file=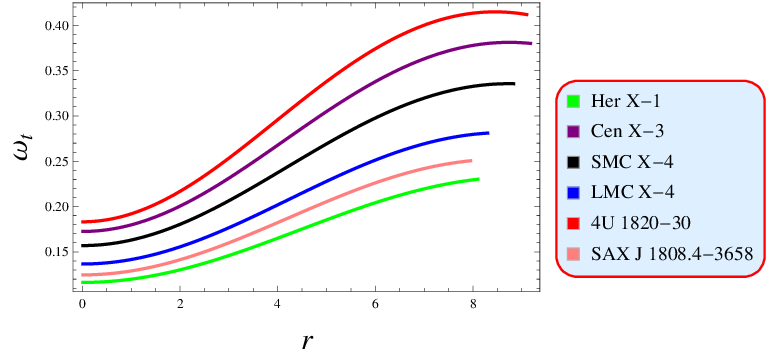,width=0.61\linewidth}
\caption{Equation of state parameters for $b=0.02$ corresponding to
model I.}
\end{figure}
\begin{figure}\center
\epsfig{file=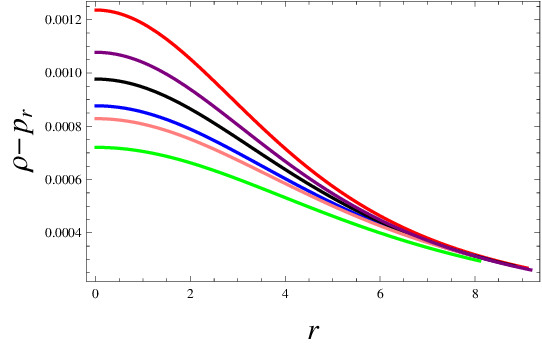,width=0.43\linewidth}\epsfig{file=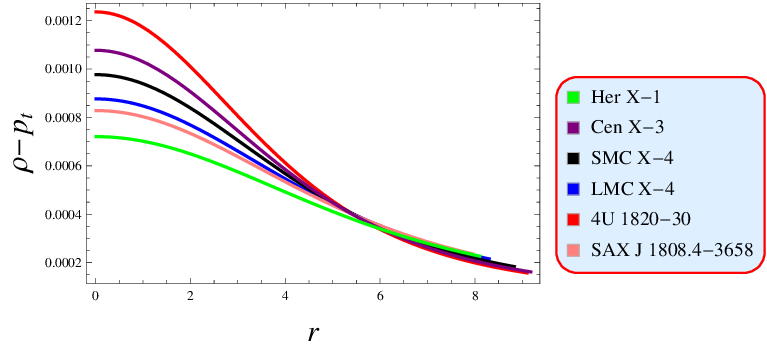,width=0.61\linewidth}
\epsfig{file=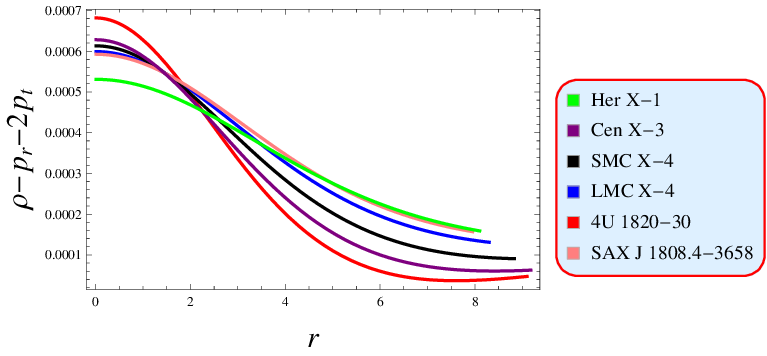,width=0.61\linewidth} \caption{ Energy
bounds for $b=0.02$ corresponding to model I.}
\end{figure}
\begin{figure}\center
\epsfig{file=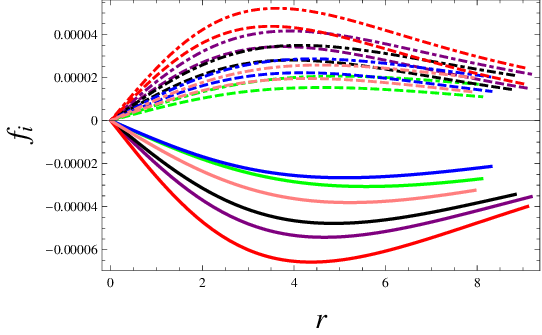,width=0.44\linewidth}\epsfig{file=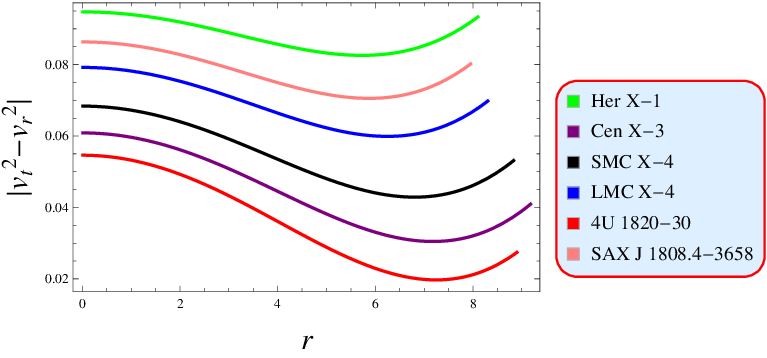,width=0.6\linewidth}
\epsfig{file=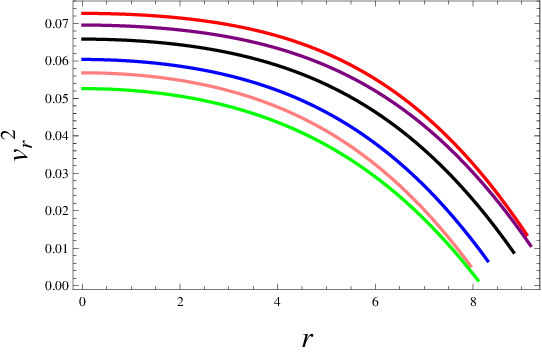,width=0.43\linewidth}\epsfig{file=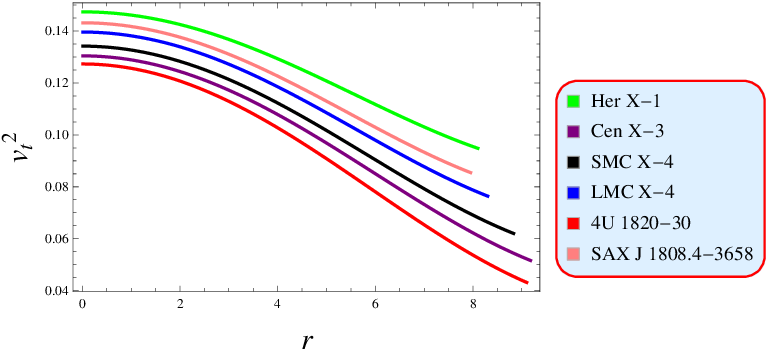,width=0.6\linewidth}
\caption{Equilibrium of forces, cracking, and sound speeds for
$b=0.02$ corresponding to model I.}
\end{figure}
\begin{figure}\center
\epsfig{file=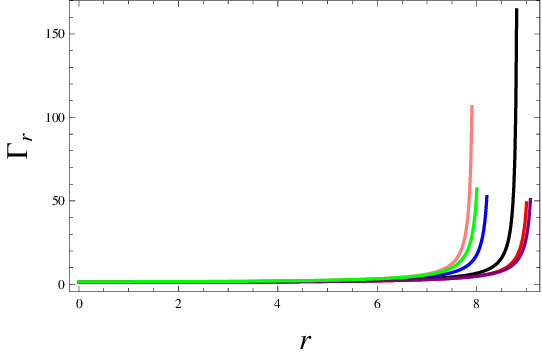,width=0.43\linewidth}\epsfig{file=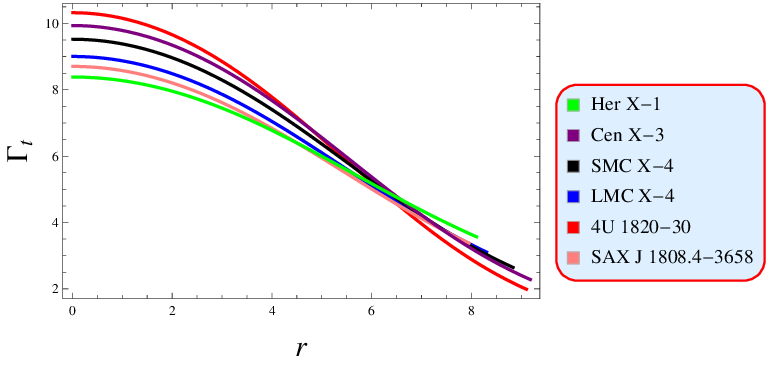,width=0.6\linewidth}
\caption{Adiabatic index for $b=0.02$ corresponding to model I.}
\end{figure}

\subsection{Model II}

Here, we employ a different additional restriction to determine the
analytic solution of the field equations in order to determine
whether or not the related results assist in modeling a compact
structure that is physically worthwhile. We take into account the
anisotropic factor and a specific $g_{rr}$ function in the preceding
analysis. The later term is expressed as \cite{42c}
\begin{align}\label{g54}
\mathrm{B}_1^2(r)=\frac{x(2r^2+y)}{(x-r^2) (r^2+y)},
\end{align}
where $x$ and $y$ are two arbitrary constants having the dimension
of $\ell^2$. The anisotropic factor \eqref{g12d}, after substituting
Eq.\eqref{g54}, becomes
\begin{align}\nonumber
\Pi(r)&=\frac{1}{8\pi r x \mathrm{A}_0(r)(2\mathrm{h}_2)^2}\big[
r^3\mathrm{A}_0(r)(2x+y)-\mathrm{A}_0'(r)\{r^4(2 x-y)\\\label{g54a}
&+4
r^2xy+xy^2\}-r(\mathrm{h}_1)(2r^4+3r^2y+y^2)\mathrm{A}_0''(r)\big],
\end{align}
which is again a differential equation of the second order.
Rearranging the equation given above results in
\begin{align}\label{g53b}
\frac{\mathrm{A}_1''}{\mathrm{A}_1}+\frac{\mathrm{A}_1'}{\mathrm{A}_1}
\bigg[\frac{r^4(2x-y)+4r^2xy+xy^2}{r(r^2-x)(2r^4+3r^2y+y^2)}\bigg]=
\frac{r\Pi(r)\big(2r^2+y^2\big)-r^3(2x+y)}{r\big(r^2-x)(2r^4+3r^2y+y^2\big)}.
\end{align}

One can find the solution of Eq.\eqref{g53b} if a specific form for
the $\Pi(r)$ is provided. We already discussed the criteria for the
anisotropic factor under which the resulting solution represents a
physically valid dense structure. Following this, the anisotropic
form that fulfills that criteria and facilitates us in analytically
solving Eq.\eqref{g53b} is
\begin{align}\label{g53f}
\Pi(r)=\frac{r^2(2x+y)}{2r^2+y^2},
\end{align}
which leads Eq.\eqref{g53b} to
\begin{align}\label{g54f}
\frac{\mathrm{A}_1''}{\mathrm{A}_1}+\frac{\mathrm{A}_1'}{\mathrm{A}_1}
\bigg[\frac{r^4(2x-y)+4r^2xy+xy^2}{r(r^2-x)(2r^4+3r^2y+y^2)}\bigg]=0,
\end{align}
through which $\mathrm{A}_1$ is analytically determined by
\begin{align}\label{g54h}
\mathrm{A}_1=\frac{D_1\sqrt{\frac{\mathrm{h}_1}{\mathrm{h}_2}}
\{\sqrt{\mathrm{h}_2}(2r^2+y)\sqrt{\frac{\mathrm{h}_1}{\mathrm{h}_2}}+\sqrt{y}
\mathrm{h}_2 \sqrt{\frac{2r^2+y}{\mathrm{h}_2}}
\mathrm{h}_3\}}{\sqrt{\mathrm{h}_1(2r^2+y)}}+D_2,
\end{align}
where $\mathrm{h}_1={r^2-x}$, $\mathrm{h}_2={r^2+y}$, and
$\mathrm{h}_3 = \mathrm{E}\{\sin ^{-1}(\frac{\sqrt{y}}{\sqrt{2}
\sqrt{r^2+y}})|\frac{2(x+y)}{y}\}$. The regularity conditions
require $\mathrm{A}_1^2(r)|_{r=0}$ to give a finite constant at
$r=0$, while its first derivative must vanish at the origin to
guarantee a non-singular gravitational potential. The plots in
Figure \textbf{9} demonstrate a physically satisfactory trend of
both $g_{tt}$ and $g_{rr}$ functions.
\begin{figure}\center
\epsfig{file=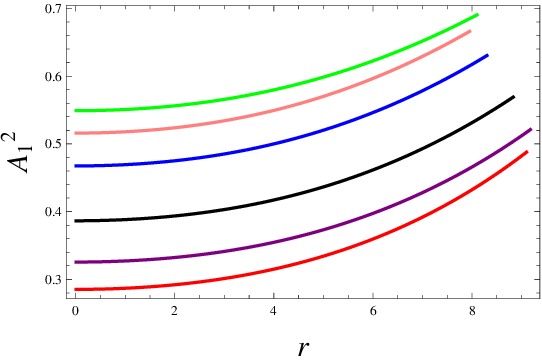,width=0.43\linewidth}\epsfig{file=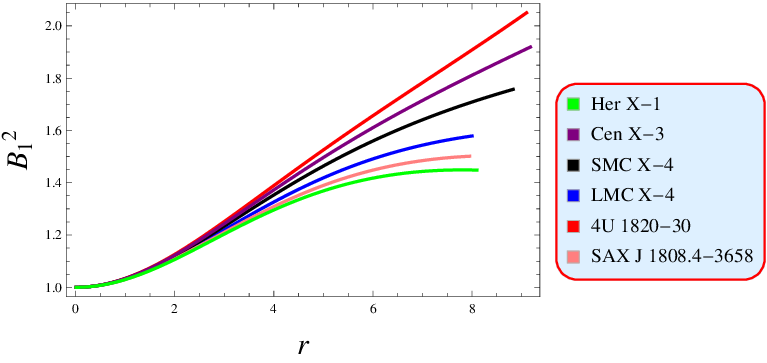,width=0.6\linewidth}
\caption{Temporal \eqref{g54h} and radial \eqref{g54} metric
functions for $y=30$ corresponding to model II.}
\end{figure}

Finally, the triplet of physical matter parameters $(\rho,p_r,p_t)$
in the form of Eqs.\eqref{g54} and \eqref{g54h} are
\begin{align}\label{g56h}
\rho &=\frac{6r^4+r^2(2x+7y)+3y(x+y)}{8 \pi x(2r^2+y)^2},
\\\nonumber
p_r &=\bigg[8 \pi x\mathrm{h}_2^{3/2}\bigg(\frac{2
r^2+y}{\mathrm{h}_2}\bigg)^{3/2}
\sqrt{\frac{\mathrm{h}_1}{\mathrm{h}_2}}\bigg\{D_2
\mathrm{h}_2\sqrt{\mathrm{h}_1(2r^2+y)}+D_1
\{\mathrm{h}_1(2r^2+y)\\\nonumber&\times\sqrt{\mathrm{h}_2}+\sqrt{y}
\sqrt{\frac{2r^2+y}{\mathrm{h}_2}}\mathrm{h}_2^2
\sqrt{\frac{\mathrm{h}_1}{\mathrm{h}_2}}\mathrm{h}_3\}\bigg\}\bigg]^{-1}
\bigg[D_1 (x-r^2)(2r^2+y)\bigg\{\mathrm{h}_2
\sqrt{\frac{2r^2+y}{\mathrm{h}_2}}\\\nonumber&\times
\sqrt{\frac{\mathrm{h}_1}{\mathrm{h}_2}} (3r^2+x+3
y)+\sqrt{y}\sqrt{\mathrm{h}_2}(\mathrm{h}_2+x)\mathrm{h}_3\bigg\}
-D_2 \sqrt{\mathrm{h}_1}\mathrm{h}_2^{3/2}
\sqrt{2\mathrm{h}_2}\sqrt{\frac{2r^2+y}{\mathrm{h}_2}}\\\label{g56hh}&\times\sqrt{\frac{\mathrm{h}_1}
{\mathrm{h}_2}}\mathrm{h}_2+x\bigg],
\\\nonumber
p_t &=\bigg[8 \pi  x
\mathrm{h}_2^{5/2}\bigg(\frac{2r^2+y}{\mathrm{h}_2}\bigg)^{5/2}
\sqrt{\frac{\mathrm{h}_1}{\mathrm{h}_2}}\bigg\{D_2
\sqrt{\mathrm{h}_1}\mathrm{h}_2
\sqrt{2r^2+y}+D_1\{\mathrm{h}_1(2r^2+y)
\\\nonumber&\times \sqrt{\mathrm{h}_2}+\sqrt{y}\sqrt{\frac{2r^2+y}{\mathrm{h}_2}}
\mathrm{h}_2^2\sqrt{\frac{\mathrm{h}_1}{\mathrm{h}_2}}\mathrm{h}_3\}\bigg\}\bigg]^{-1}\bigg
[D_1(x-r^2)(2r^2+y)\bigg\{\mathrm{h}_2\sqrt{\frac{2r^2+y}{\mathrm{h}_2}}
\\\nonumber&\times\sqrt{\frac{\mathrm{h}_1}{\mathrm{h}_2}}\{6 r^4+8r^2 y+y(x+3
y)\}+\sqrt{y} \sqrt{\mathrm{h}_2}\{2
r^4+2r^2y+y(x+y)\}\mathrm{h}_3\bigg\}\\\label{g56hhh}&-D_2\sqrt{\mathrm{h}_1}
(\mathrm{h}_2)^{3/2} \sqrt{2r^2+y}\sqrt{\frac{2r^2+y}{\mathrm{h}_2}}
\sqrt{\frac{\mathrm{h}_1}{\mathrm{h}_2}}\{2r^4+2 r^2 y+y
(x+y)\}\bigg].
\end{align}

It is observed that the above field equations contain three
constants that must be calculated via junction conditions. We
already discussed that the exterior geometry in this context is
considered as the the Schwarzschild metric \eqref{g15}. Matching our
interior model with this exterior spacetime under the first
fundamental form gives
\begin{align}\label{g16ab}
\frac{D_1\sqrt{\frac{\mathrm{h}_1^*}{\mathrm{h}_2^*}}
\{\sqrt{\mathrm{h}_2^*}(2R^2+y)\sqrt{\frac{\mathrm{h}_1^*}{\mathrm{h}_2^*}}+\mathrm{h}_2^*\sqrt{y}
\sqrt{\frac{2R^2+y}{\mathrm{h}_2^*}}
\mathrm{h}_3^*}{\sqrt{\mathrm{h}_1^*(2R^2+y)}}+D_2
&~=~\sqrt{1-\frac{2\mathrm{M}}{R}},\\\label{g17ab}
\frac{x(2R^2+y)}{(x-R^2) (R^2+y)}&~=~1-\frac{2\mathrm{M}}{R},
\end{align}
where $\mathrm{h}_1^*=\mathrm{h}_1(R)$,
$\mathrm{h}_2^*=\mathrm{h}_2(R)$, and $\mathrm{h}_3^*=
\mathrm{h}_2(R)$. Moreover, the radial pressure is found to vanish
at a specific value of $r$, which results after merging with
Eq.\eqref{g56hh} as
\begin{align}\nonumber
p_r(R)&=2D_1\mathrm{h}_2^*(x-R^2)
2\mathrm{h}_2^*\bigg\{\mathrm{h}_2^*\sqrt{\frac{2R^2+y}{\mathrm{h}_2^*}}\sqrt{\frac{\mathrm{h}_1^*}
{\mathrm{h}_2^*}}(3R^2+x+3y)\\\label{g55h}&+\sqrt{y}
\mathrm{h}_2^{*3/2}+x\mathrm{h}_3^*\bigg\}
-D_2\mathrm{h}_2^{*3/2}\sqrt{2\mathrm{h}_1^*\mathrm{h}_2^*}
\sqrt{\frac{2\mathrm{h}_2^*}{\mathrm{h}_2^*}}\sqrt{\frac{\mathrm{h}_1^*}{\mathrm{h}_2^*}}
(\mathrm{h}_2^*+x)=0.
\end{align}
The simultaneous solution of Eqs.\eqref{g16ab}-\eqref{g55h} gives
the three constants as follows
\begin{align}\label{gg50h}
x&=\frac{R^3(R^2+y)}{4\mathrm{M}R^2+2\mathrm{M} y-R^3},
\\\label{g50hh}
D_1&=-\frac{\sqrt{1-\frac{2\mathrm{M}}{R}}(\mathrm{h}_2^*+x)}{2
\sqrt{2yR^2\mathrm{h}_1^*\mathrm{h}_2^*}},\\\nonumber
D_2&=\frac{1}{4yR^2\mathrm{h}_1^*\mathrm{h}_2^*}\sqrt{1-\frac{2\mathrm{M}}{R}}
\bigg\{2yR^2\mathrm{h}_1^*(3R^2+x+3y)\\\label{gg50hh}& +\sqrt{y}
\mathrm{h}_2^{*3/2}\sqrt{\frac{R^2}{\mathrm{h}_2^*}+1}
\sqrt{\frac{\mathrm{h}_1^*}{\mathrm{h}_2^*}}(\mathrm{h}_2^*+x)h_3^*\bigg\}.
\end{align}
The matter triplet \eqref{g56h}-\eqref{g56hhh} and its physical
properties are now graphically investigated. The profile of the
pressure components, energy density and anisotropic factor is shown
in Figure \textbf{10}. We also observe the vanishing of the radial
pressure at the interface. At the center, $p_r(0)=constant=p+t(0)$,
implying that the radial and tangential pressures are equal,
resulting in the vanishing anisotropy at that point. Additionally,
Figure \textbf{11} plots the regularity criteria for the above
matter variables, showing a required trend.
\begin{figure}\center
\epsfig{file=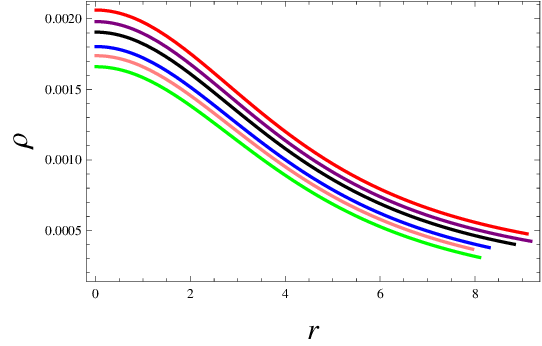,width=0.43\linewidth}\epsfig{file=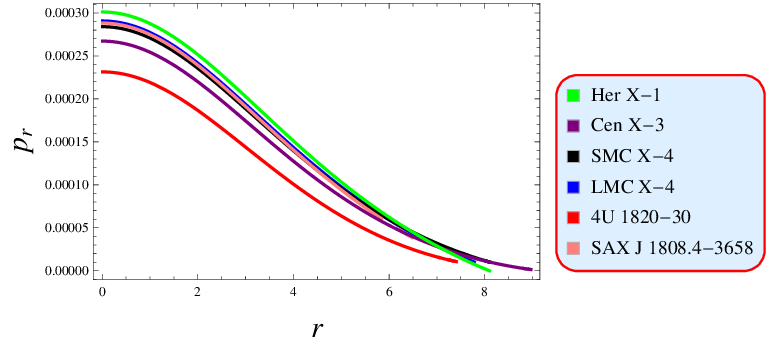,width=0.61\linewidth}
\epsfig{file=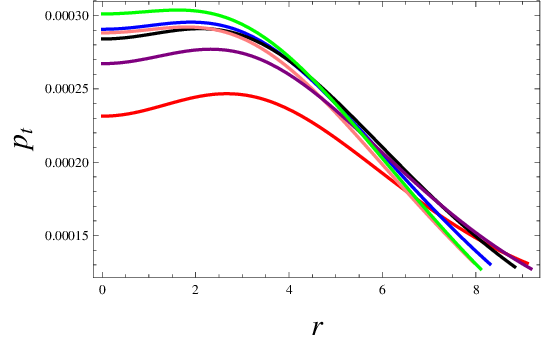,width=0.43\linewidth}\epsfig{file=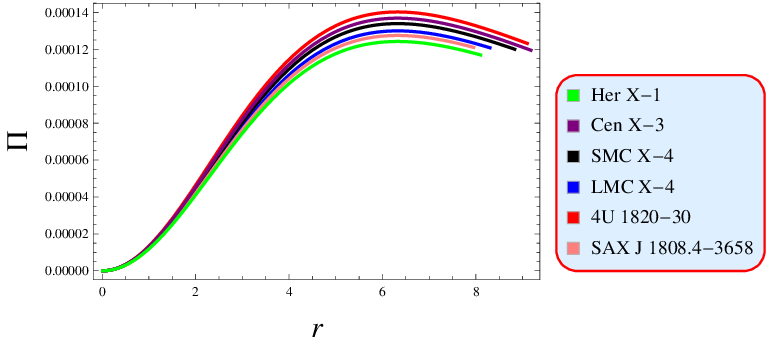,width=0.61\linewidth}
\caption{Fluid parameters for $y=30$ corresponding to model II.}
\end{figure}
\begin{figure}\center
\epsfig{file=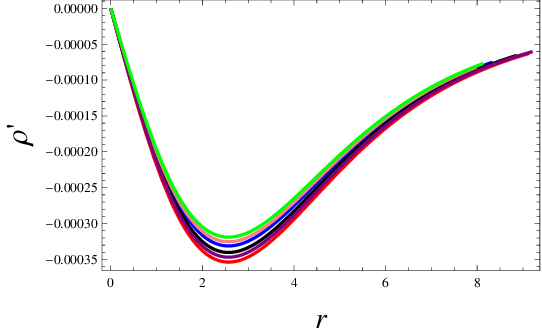,width=0.43\linewidth}\epsfig{file=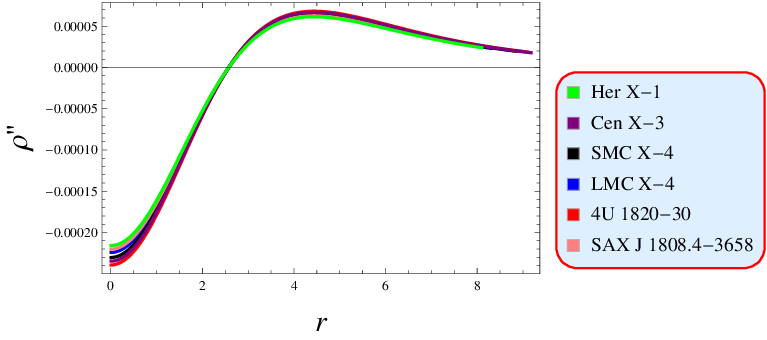,width=0.6\linewidth}
\epsfig{file=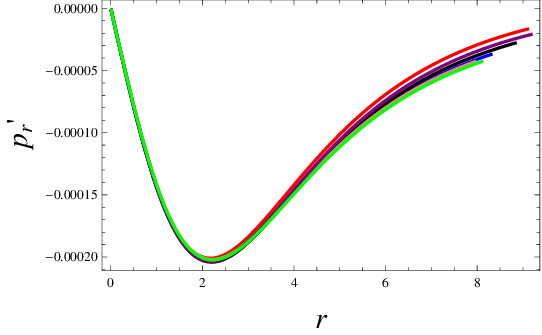,width=0.43\linewidth}\epsfig{file=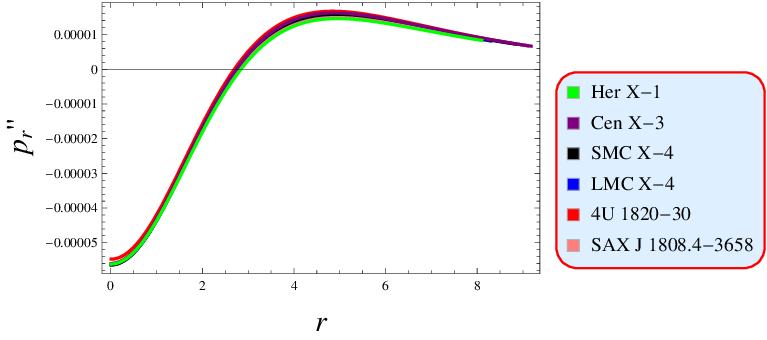,width=0.61\linewidth}
\epsfig{file=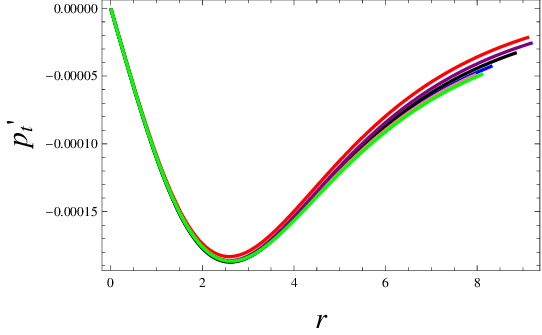,width=0.43\linewidth}\epsfig{file=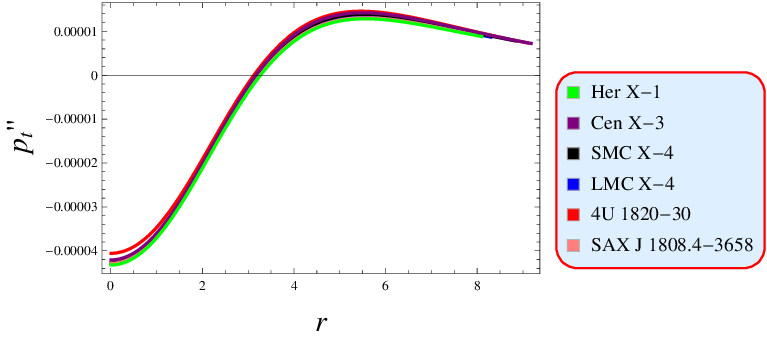,width=0.61\linewidth}
\caption{Regularity conditions for $y=30$ corresponding to model
II.}
\end{figure}
\begin{table}[H]
\scriptsize \centering \caption{Various physical factors for
different stars corresponding to model II.} \label{Table3}
\vspace{+0.1in} \setlength{\tabcolsep}{1.55em}
\begin{tabular}{cccccc}
\hline\hline \textbf{Stars} & $\rho_c {\bf (gm/cm^3)}$ & $\rho_s
{\bf (gm/cm^3)}$ & $p_{c} {\bf (dyne/cm^2)}$
\\\hline $\mathrm{Her X-1}$ & 2.2181$\times$10$^{15}$ & 3.9439$\times$10$^{14}$ &
3.6024$\times$10$^{35}$
\\\hline
$\mathrm{Cen X-3}$ & 2.6422$\times$10$^{15}$ &
5.5788$\times$10$^{14}$ & 3.1960$\times$10$^{35}$
\\\hline
$\mathrm{SMC X-4}$ & 2.5379$\times$10$^{15}$ &
5.2457$\times$10$^{14}$ & 3.4089$\times$10$^{35}$
\\\hline $\mathrm{LMC X-4}$ & 2.4028$\times$10$^{15}$ & 4.8577$\times$10$^{14}$ &
3.4870$\times$10$^{35}$
\\\hline
$\mathrm{4U 1820-30}$ & 2.7479$\times$10$^{15}$ &
6.3320$\times$10$^{14}$ & 2.7692$\times$10$^{35}$
\\\hline
$\mathrm{SAX J 1808.4-3658}$ & 2.3091$\times$10$^{15}$ &
4.7172$\times$10$^{14}$ &
3.4461$\times$10$^{35}$ \\
\hline\hline
\end{tabular}
\end{table}
\begin{figure}\center
\epsfig{file=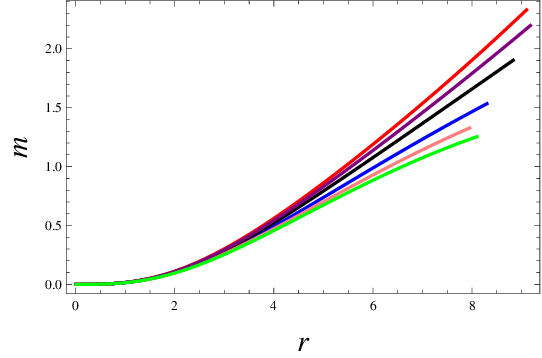,width=0.43\linewidth}\epsfig{file=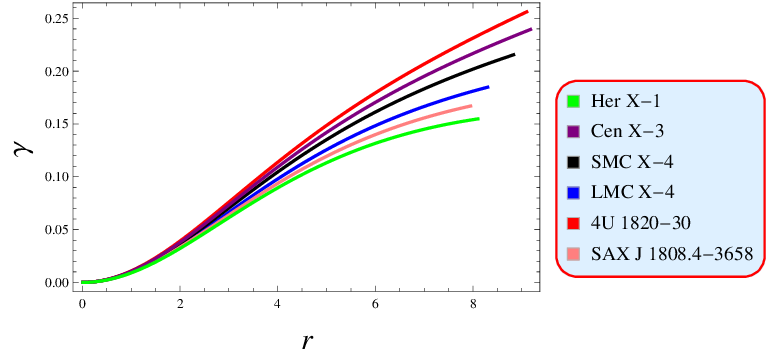,width=0.61\linewidth}
\epsfig{file=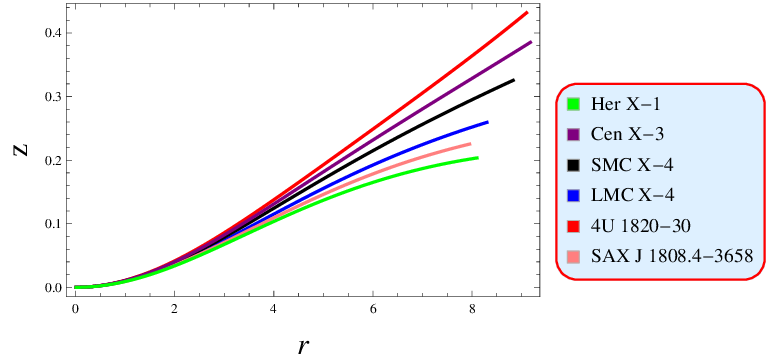,width=0.61\linewidth} \caption{Physical
factors for $y=30$ corresponding to model II.}
\end{figure}

Table \textbf{3} presents the central density, surface density and
central pressure for all the considered starts under our model II.
We notice that their values lie within the acceptable ranges which
ultimately points the existence of dense astrophysical objects. The
mass function, redshift, and compactness factors are shown in Figure
\textbf{12}, from which we see that the values of mass corresponding
to all the considered stars are consistent with their observed
masses. The behavior of the equation of state parameters is
illustrated in Figure \textbf{13} that confirms the physical
validity of our model II. We display only the trace and dominant
energy constraints in Figure \textbf{14} which suggests that the
suggested solution is viable and thus the corresponding interior
contains the ordinary fluid. The net sum of all forces is found to
be zero in Figure \textbf{15}, hence, the model II is in the state
of equilibrium. We further analyze the stability in the same Figure
using cracking and sound speed criteria, which confirms the
stability of model II over the whole interior. The stability
analysis is also performed through the adiabatic index in Figure
\textbf{16}, and we observe the required results which are
consistent with the earlier two tests.
\begin{figure}\center
\epsfig{file=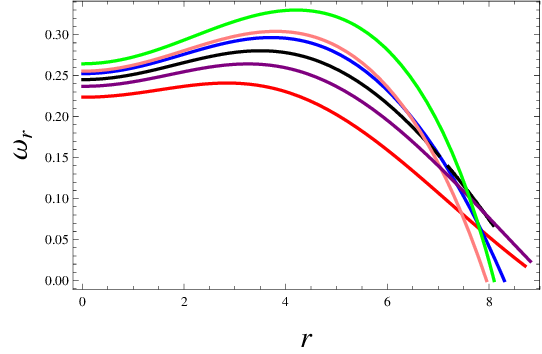,width=0.43\linewidth}\epsfig{file=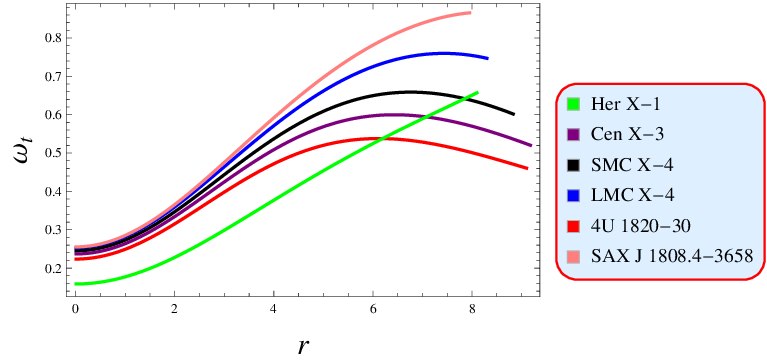,width=0.6\linewidth}
\caption{Equation of state parameters for $y=30$ corresponding to
model II.}
\end{figure}
\begin{figure}\center
\epsfig{file=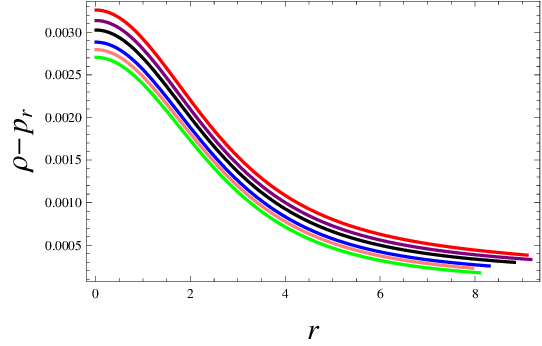,width=0.43\linewidth}\epsfig{file=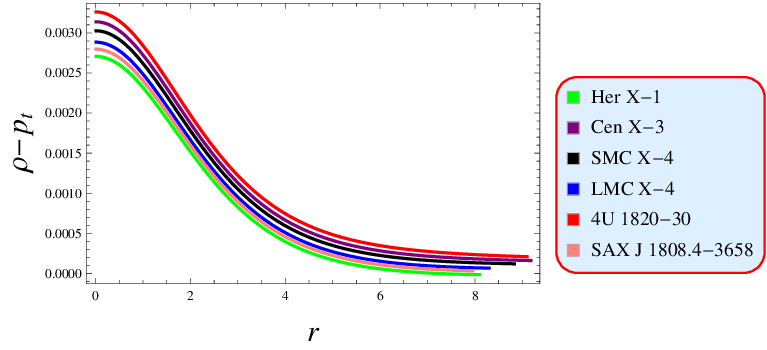,width=0.6\linewidth}
\epsfig{file=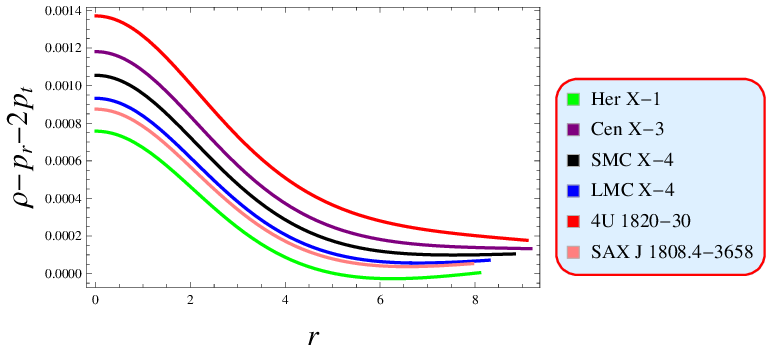,width=0.6\linewidth} \caption{Energy
bounds for $y=30$ corresponding to model II.}
\end{figure}
\begin{figure}\center
\epsfig{file=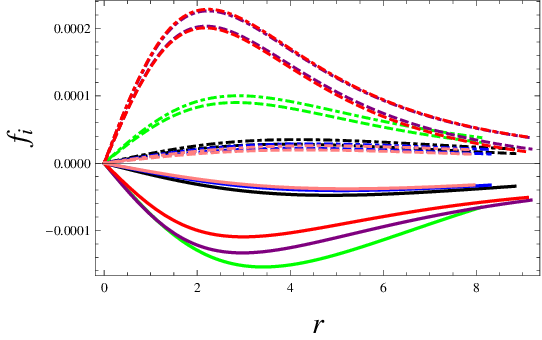,width=0.44\linewidth}\epsfig{file=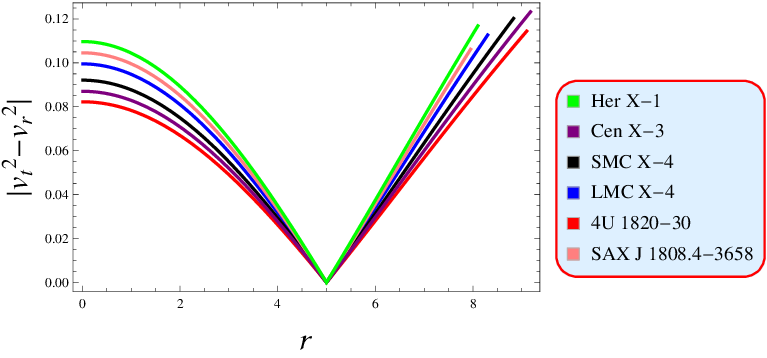,width=0.6\linewidth}
\epsfig{file=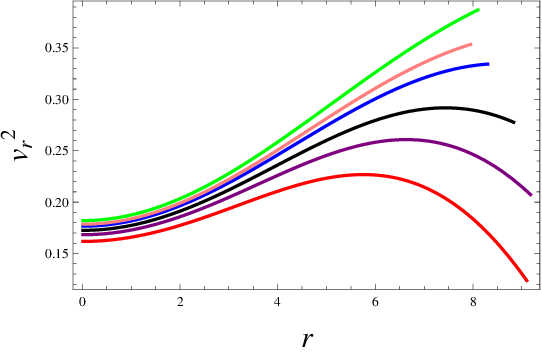,width=0.43\linewidth}\epsfig{file=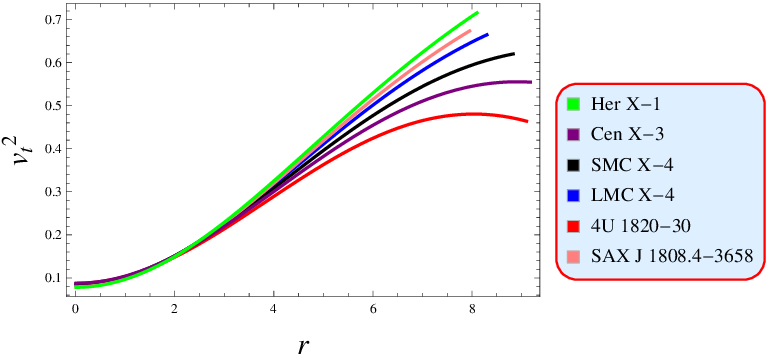,width=0.6\linewidth}
\caption{Equilibrium of forces, cracking, and sound speeds for
$y=30$ corresponding to model II.}
\end{figure}
\begin{figure}\center
\epsfig{file=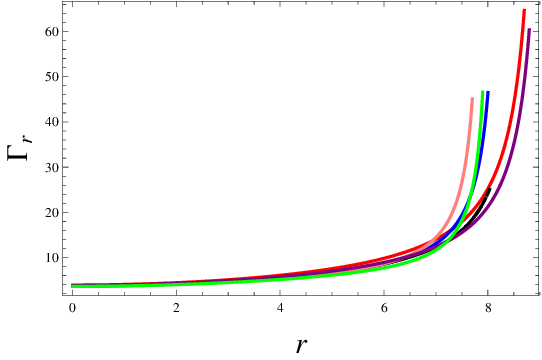,width=0.43\linewidth}\epsfig{file=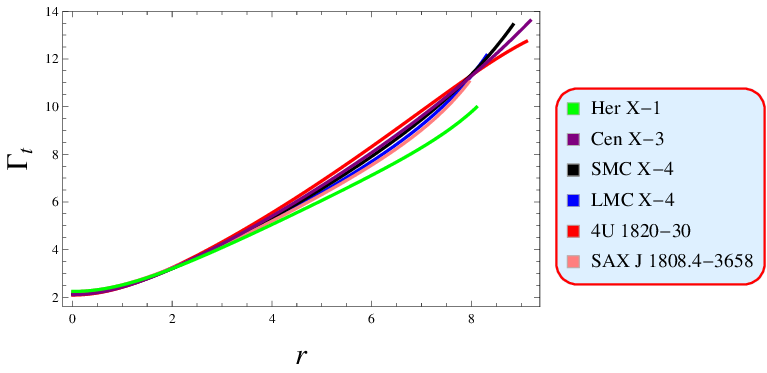,width=0.6\linewidth}
\caption{Adiabatic index for $y=30$ corresponding to model II.}
\end{figure}

\section{Conclusions}

Two novel singularity-free anisotropic solutions are developed in
this study through the utilization of different constraints. To
achieve this, we have taken a static spherical interior spacetime
and the corresponding governing field equations as well as the
gravitational mass function have been constructed. Given that the
system was under-determined, specific constraints have been
introduced to obtain exact analytical solutions. We have considered
certain types of $g_{rr}$ metric potential and merged it with the
anisotropy which resulted in differential equations of second-order
involving the $g_{tt}$ function. The temporal metric functions have
then been determined by assuming particular types of the
anisotropies (null in the core and increasing outwards), which
results in our models I and II. At the spherical boundary, the
constants associated with these metric functions have been derived
by matching the interior and the external Schwarzschild spacetimes.
Below is the detailed summary of the graphical analysis conducted
for both our models for six distinct compact stars (Table
\textbf{1}).
\begin{itemize}
\item We have determined that the metric functions and related matter
determinants of both our models show the an appropriate profile
which is required for the physical existence of stellar models
(Figures \textbf{1, 2, 9} and \textbf{10}). At the spherical
boundary, the radial pressures \eqref{g55i} and \eqref{g56hh} reduce
to zero which is also consistent with the second fundamental form of
the boundary conditions. Tables \textbf{2} and \textbf{3} present
the numerically computed values of these fluid parameters which lie
in the acceptable range for stars to exist both theoretical and
physical point of view.

\item The anisotropy shows an increasing trend in both models that
suggests a repulsive effect. Hence, our models remain stable for a
long period of time. Additionally, the regularity conditions are
also satisfied, as illustrated in Figures \textbf{3} and
\textbf{11}. The results for the redshift, compactness, and mass
functions are plotted in Figures \textbf{4} and \textbf{12} for
models I and II, respectively. We observe that their profiles are
consistent with the observations.

\item Figures \textbf{5} and \textbf{13} clearly demonstrate that both
parameters $\omega_r$ and $\omega_t$ are less than one that confirm
the validity of our models. The physical viability of both resulting
models is checked in Figures \textbf{6} and \textbf{14}, where the
corresponding energy bounds are fully satisfied which ultimately
supports the presence of normal fluid within the resulting
interiors.

\item The balance of all the forces acting on a self-gravitating bodies
is also important to be checked. We have performed this analysis in
Figures \textbf{7} and \textbf{15}, and deduced the hydrostatic
equilibrium of our models as the sum of all the acting forces
nullifies. In the same Figure, we have used two different tests,
sound speed fluctuations and cracking criteria, to explore the
stability. We have found that both criteria are fulfilled, claiming
models I and II to be stable. Moreover, the adiabatic index is
checked that also supports the stability (Figures \textbf{8} and
\textbf{16}).

\item It is concluded that both of our models satisfy a family of
stability requirements and physical characteristics necessary for
celestial bodies' existence in any gravitational theory. We,
therefore, claim that the developed results are effective for
modeling compact interiors coupled with the anisotropic fluid
distribution.
\end{itemize}

The implications of this study extend to both theoretical and
observational astrophysics. Theoretically, it reinforces the
necessity of incorporating anisotropy in modeling compact stars, as
isotropic assumptions often fail to capture the complex dynamics of
high-density matter. Observationally, the models align with
empirical data from known compact stars, offering a framework to
interpret their mass-radius relationships and internal structures.
Future work could explore the inclusion of more sophisticated
equations of state, magnetic fields, or rotational effects to
further refine these models. Additionally, extending this analysis
to modified gravity theories or higher-dimensional spacetimes may
uncover new phenomena in extreme gravitational environments.
Ultimately, this study lays a solid foundation for future
investigations into the enigmatic nature of compact objects and
their role in the cosmos.
\\\\
\textbf{Data Availability Statement:} No new data is generated in
this study.


\begin{thebibliography}{00}
\bibitem{1} K. Schwarzschild, Sitz. Deut. Akad. Wiss Berlin Kl. Math. Phys. \textbf{1916}, 189
(1916)

\bibitem{2} K. Schwarzschild, Sitz. Deut. Akad. Wiss Berlin Kl. Math. Phys. \textbf{24}, 424
(1916)

\bibitem{56} L. Herrera, N.O. Santos, Phys. Rep. \textbf{286}, 53 (1997)

\bibitem{57} J. Ovalle, Phys. Rev. D \textbf{95}, 104019 (2017)

\bibitem{58} J. Ovalle, R. Casadio, R. da Rocha, A. Sotomayor, Eur. Phys. J. C \textbf{78}, 122 (2018)

\bibitem{57a} T. Naseer, Phys. Dark Universe
\textbf{46}, 101663 (2024)

\bibitem{57b} T. Naseer, M. Sharif, S. Manzoor, A. Fatima, Mod. Phys. Lett. A
\textbf{39}, 2450048 (2024)

\bibitem{64} J. Ospino, L.A. Nunez, Eur. Phys. J. C \textbf{80}, 166 (2020)

\bibitem{66} B.V. Ivanov, Eur. Phys. J. C \textbf{81}, 227 (2021)

\bibitem{66a} Y. Feng et al., Chin. J. Phys. \textbf{90}, 372-386 (2024)

\bibitem{67} T. Naseer, J.L. Said, Eur. Phys. J. C \textbf{84}, 808 (2024)

\bibitem{68} L. Herrera, A. Di Prisco, J. Ospino, E. Fuenmayor,
J. Math. Phys. \textbf{42}, 2129 (2001)

\bibitem{79} W. Baade, F. Zwicky, Phys. Rev. \textbf{46}, 76 (1934)

\bibitem{3} J. Jeans, Mon. Not. R. Astron. Soc. \textbf{82}, 122 (1922)

\bibitem{4} G. Lemaitre, Ann. Soc. Sci. Brux. A \textbf{53}, 51 (1933)

\bibitem{5} M. Ruderman, Annu. Rev. Astron. Astrophys. \textbf{10}, 427 (1972)

\bibitem{10} R.F. Sawyer, Phys. Rev. Lett. \textbf{29}, 382 (1972)

\bibitem{6} S.S. Yazadjiev, Phys. Rev. D \textbf{85}, 044030 (2012)

\bibitem{7} C.Y. Cardall, M. Prakash, J.M. Lattimer, Astrophys. J. \textbf{554}, 322 (2001)

\bibitem{8} R. Ciolfi, V. Ferrari, L. Gualtieri, Mon. Not. R. Astron. Soc. \textbf{406}, 2540 (2010)

\bibitem{9} J. Frieben, L. Rezzolla, Mon. Not. R. Astron. Soc. \textbf{427}, 3406 (2012)

\bibitem{11} V. Canuto, Annu. Rev. Astron. Astrophys. \textbf{12}, 167 (1974)

\bibitem{12} H. Heiselberg, M. Hjorth-Jensen, Phys. Rep. \textbf{328}, 237 (2000)

\bibitem{13} R.L. Bowers, E.P.T. Liang, Astrophys. J. \textbf{188}, 657 (1974)

\bibitem{14} Z. Roupas, Astrophys. Space Sci. \textbf{366}, 9 (2021)

\bibitem{15} D. Deb, B. Mukhopadhyay, F. Weber, Astrophys. J. \textbf{922}, 149 (2021)

\bibitem{16} L. Herrera, J. Ospino, A. Di Prisco, Phys. Rev. D \textbf{77},
027502 (2008)

\bibitem{17} L. Herrera, W. Barreto, Phys. Rev. D \textbf{88},
084022 (2013)

\bibitem{18} S. Das, B.K. Parida, S. Ray, S.K. Pal, Phys. Sci. Forum \textbf{2}, 29 (2021)

\bibitem{19} K. Lake, Phys. Rev. D \textbf{80}, 064039 (2009)

\bibitem{20} L. Herrera, N.O. Santos, Phys. Rep. \textbf{286}, 53 (1997)

\bibitem{21} L. Herrera, Phys. Rev. D \textbf{101}, 104024 (2020)

\bibitem{47} M.K. Gokhroo, A.L. Mehra, Gen. Relativ. Gravit.
\textbf{26}, 75 (1994)

\bibitem{30} N. Pant, R.N. Mehta, M. Pant, Astrophys. Space Sci.
\textbf{332}, 473 (2011)

\bibitem{31} Y.K. Gupta, S.K. Maurya, Astrophys. Space Sci.
\textbf{332}, 155 (2011)

\bibitem{36} A. Rehman et al., Nucl. Phys. B \textbf{1013}, 116852 (2025)

\bibitem{21d} S.K. Maurya, Y.K. Gupta, S.
Ray, B. Dayanandan, Eur. Phys. J. C \textbf{75}, 225 (2015)

\bibitem{21e} S.K. Maurya, A. Banerjee, S. Hansraj, Phys. Rev. D
\textbf{97}, 044022 (2018)

\bibitem{21g} S.K. Maurya, S.D. Maharaj, J. Kumar, A.K. Prasad, Gen. Relativ. Gravit. \textbf{51},
86 (2019)

\bibitem{21h} A. Rehman et al., Nucl. Phys. B \textbf{1015}, 116897 (2025)

\bibitem{33} M.F. Shamir, G. Mustafa, Q. Hanif, Int. J. Mod. Phys. A
\textbf{35}, 2050083 (2020)

\bibitem{34} K. Hassan et al., Chin. J. Phys. \textbf{91}, 916 (2024)

\bibitem{46a} B. Siza et al., Eur. Phys. J. C \textbf{84},
1203 (2024)

\bibitem{46b} T. Naseer, Chin. J. Phys. \textbf{96}, 1212-1231 (2025)

\bibitem{46c} M. Al Hadhrami et al., Pramana \textbf{97}, 13 (2022)

\bibitem{46d} S.K. Maurya et al., Mon. Not. R. Astron. Soc.
\textbf{519}, 4303 (2023)

\bibitem{46e} T. Naseer, Eur. Phys. J. C \textbf{84}, 1256 (2024)

\bibitem{40} L. Baskey, S. Ray, S. Das, S. Majumder, A. Das, Eur. Phys. J. C \textbf{83}, 307 (2023)

\bibitem{37} M. Sharif, T. Naseer, Int. J. Mod. Phys. D \textbf{31}, 2240017 (2022)

\bibitem{38} M. Sharif, Q. Ama-Tul-Mughani, Chin. J. Phys. \textbf{65}, 207 (2020)

\bibitem{fag} T. Naseer, Astropart. Phys. \textbf{166}, 103073 (2025)

\bibitem{fah} S. Sarkar, N. Sarkar, P. Rudra, F. Rahaman, T.
Ghorui, Eur. Phys. J. C \textbf{83}, 1005 (2023)

\bibitem{fak} E. Demir, et al., Chin. J. Phys.
\textbf{91}, 299 (2024)

\bibitem{faj} C.W. Misner, D.H. Sharp, Phys. Rev. \textbf{136}, B571 (1964)

\bibitem{ab} M.S.R. Delgaty, K. Lake, Comput. Phys. Commun. \textbf{115}, 395 (1998)

\bibitem{ac} T. Naseer, Ann. Phys. \textbf{479}, 170035 (2025)

\bibitem{42a} H.A. Buchdahl, Phys. Rev. \textbf{116}, 1027 (1959)

\bibitem{42b} B.V. Ivanov, Phys. Rev. D \textbf{65}, 104011 (2002)

\bibitem{42c} R.C. Tolman, Phys. Rev. \textbf{55}, 364 (1939)

\bibitem{42d} J.R. Oppenheimer, G.M. Volkoff, Phys. Rev. \textbf{55}, 374 (1939)

\bibitem{42bb} H. Abreu, H. Hernandez, L.A. Nunez, Class. Quantum Grav.
\textbf{24}, 4631 (2007)

\bibitem{42ba} L. Herrera, Phys. Lett. A \textbf{165}, 206 (1992)

\bibitem{42f} H. Heintzmann, W. Hillebrandt, Astron. Astrophys.
\textbf{38}, 51 (1975)

\bibitem{88a} M. Kohler, K.L. Chao, Z. Naturforsch. A
\textbf{20}, 1537 (1965)

\bibitem{88b} R.S. Tikekar, Curr. Sci. \textbf{39}, 460 (1970)
\end{thebibliography}
\end{document}